\documentclass[prd,twocolumn,showpacs,amssymb,superscriptaddress]{revtex4}
\usepackage{latexsym}
\usepackage{dcolumn}
\usepackage{graphicx}
\usepackage{latexsym}
\usepackage{subfigure}
\usepackage{longtable}
\usepackage{amsmath}
\usepackage{supertabular}
\usepackage{ccaption}
\usepackage{natbib}
\usepackage[latin1]{inputenc}
\usepackage[thinspace]{SIunits}
\addunit{\parsec}{pc}
\addunit{\ergios}{erg}
\addunit{\anho}{yr}
\addunit{\edad}{\giga\anho}
\def \cm {\centi\meter}
\def \be {\begin{equation}}
\def \ee {\end{equation}}

\def \phu {\kilo\meter\,\reciprocal\second\mega\reciprocal\parsec}
\def \kev {\kilo\electronvolt}
\def \upb {\ergios\ \rpsquare\cm\,\reciprocal\second}
\def \usb {\ergios\ \rpsquare\cm}
\def \dt {\gram\,\rpcubic\cm}
\def \s   {\second}
\def \bd  {$\ldots$}
\def \cd  {$\cdots$}
\def \sr {$\pm$}
\def \om {$\Omega_{\textrm{m}}$}
\def \omm {\Omega_{\textrm{m}}}

\def \oxx  {\Omega_{\textrm{x}}}
\def \siom {$\sigma_{\Omega_{m}}$}

\def \si {$\sigma_{n}$}
\def \ho {$H_0$}
\def \epk {$E_{\textrm{peak}}$}
\def \epkk {E_{\textrm{peak}}}
\def \pbo {$P_{\textrm{bolo}}$}
\def \sbo {$S_{\textrm{bolo}}$}

\def \rcmb {${\cal R}$}
\def \bao {${\cal A}$}
\def \chidof {$\chi^2_{\text{dof}}$}
\def \zc {$z_{\text{card}}$}
\def \zac {$z_{\text{acc}}$}
\def \te {\theta}
\def \sg {\sigma}
\def \m {\text{m}}
\def \trt {\tau_{\text{\tiny RT}}}
\def \tlag {\tau_{\text{\tiny lag}}}
\def \prh {Prior $H_0$}
\begin{document}

\pretolerance 10000

\title{Confronting the Hubble Diagram of Gamma-Ray Bursts with Cardassian Cosmology}

\author{Herman J.~Mosquera Cuesta$^{1,\ast}$, Habib Dumet M.$^{1,\ast}$ and Cristina Furlanetto }
\email{hermanjc@cbpf.br :::: hdumetm@cbpf.br :::: crisf@cbpf.br}
\affiliation{\mbox{Instituto de Cosmologia, Relatividade e Astrof\'{\i}sica (ICRA-BR), Centro Brasileiro
de Pesquisas F\'{\i}sicas } \\ \mbox{Rua Dr. Xavier Sigaud 150, Cep 22290-180, Urca, Rio de Janeiro, RJ,
Brasil } }

\begin{abstract}
We construct the Hubble diagram of Gamma-Ray Bursts (GRBs) with
redshifts reaching up to $z \sim 6$, by  using five luminosity vs.
luminosity indicator relations calibrated with the Cardassian
cosmology. This model has a major interesting feature: despite of 
being matter-dominated and flat, it can explain the present
accelerate expansion of the universe. This is the first study of
this class of models using high redshift GRBs. We have performed a
$\chi$-square statistical analysis of the GRBs calibrated with the
Cardassian model, and also combined them with both the current
Cosmic Microwave Background (CMB) and Baryonic Acoustic
Oscillation (BAO) data. Our results show consistency between the
current observational data and the model predictions, in
particular, the best-fit parameters obtained from that
$\chi^2$-analysis are in agreement with those obtained from
previous investigations. The influence of these best-fit
parameters on the redshift at which the universe would start to
follow the Cardassian expansion, i. e.,  \zc,\, and on both the
redshift at which the universe supposedly had started to accelerate, 
i. e., \zac,\, 
and the age-redshift relation $H_0t_0$, are also discussed. Our results 
also show that the universe, from the point of view of GRBs, had undergo 
a transition to acceleration at a redshift $z \approx 0.2-0.7$, which 
agrees with the SNIa results. One important point that we notice is that 
despite the statistical analysis is performed with a model that does not 
need any vaccum energy, we found that the results attained using this 
cosmological model are compatible with those obtained with the concordance 
cosmology ($\Lambda$-CDM), as far as GRBs is concerned. Hence, after 
confronting the Cardassian scenario against the GRBs HD, our main 
conclusion is that GRBs should indeed be considered a complementary 
tool to several other observational studies for doing precision cosmology.
\end{abstract}

\date{\today}


\maketitle

\section{Introduction}

The discovery of the present late-time acceleration of the
universe with observations of Type Ia Supernovae (SNIa)
\cite{riess98,perl99}, corroborated with the cosmic microwave
background \cite{benett2003} and the large scale structure 
observations \cite{tegmark2003}, have motivated the introduction 
of several cosmological scenarios. Among them one can quote the 
model with a positive cosmological constant where the dark energy 
evolves with  time \cite{linder2003}, the model with an exotic 
equation state known
as the Generalized Chaplygin Gas (GCG), which has the interesting
feature of allowing the unification of dark energy and dark matter
\cite{ugo2001,bento2002}, and similar scenarios. Yet, another possible
explanation for this accelerate expansion could be the infrared
modification of gravity as predicted by extra-dimension physics,
which would lead to a modification of the effective Friedmann's
equation at late times. Such modifications may arise as a
consequence of our 4-dimensional universe being a surface, or a
brane, embedded into a higher dimensional bulk space-time to which
only gravity could spread\cite{extd,rs,dgp,chfr}. Alternatively,
they may arise if there is dark matter self-interactions
characterized by negative pressure\cite{gondo}.

One of these interesting possibilities is the modification of the
Friedmann's equation by the introduction of an additional nonlinear
term of mass. Such an idea is referred to as the Cardassian model\cite{free}.

\indent On the other hand, although SNIa are at the base of the suggestion of 
the late-time acceleration of the universe, the fact is that up to the present 
time it has not been possible to register any supernova event at redshift $z>2$.
Therefore, if we wish to know the actual expansion history of our universe
we need to trace it back over a wide range of redshifts. In this perspective, the
discovery of the X-ray afterglow of the gamma-ray burst (GRB) event in 1997 february
28, by the Beppo-SAX sattelite \cite{costa}, allowed the first precise measurement
of the redshift of a GRB. This breakthrough definitely confirmed the long-standing
suspicion that GRBs might have cosmological origin. That event also makes it possible
to use GRBs as actual probes in contemporary astrophysics and cosmology. This new
window onto the universe has the advantage of allowing to follow well back in time,
up to very high redshifts, the expansion history of our universe.

\indent In fact, the possibility of using GRBs as cosmological
probes has estimulated the search for self-consistent methods of
bringing them into the realm of cosmology. Those procedures include
the Amati relation \cite{amati1}, the Ghirlanda relation \cite{ghirlanda2004a},
the Liang \& Zhang relation \cite{liang2005}, and the Firmani relation
\cite{firmani2006,lazzati}. All of these relations take into account the most
relevant physical properties of GRBs as the peak  energy, jet openning
angle, time lag and variability. In these lines, recently, Schaefer
\cite{bs2007} (hereafter BS2007) used five luminosity vs. luminosity
indicator relations to calibrate GRBs for a specific cosmology. The use
of those relations turns the GRBs reliable standard candles for
practical studies in precision cosmology. A similar technique has 
been implemented by Mosquera Cuesta et al. in Ref.\cite{nos2007}. 

As is well-known, in the case of supernovae the calibration does not
depend on any cosmology because of a large set of those SNIa are nearby
events. In practice, however, many of the supernovae in Hamuy et al.
sample \cite{hamuy} are definitely distant so that the effect of a varying
cosmology must be introduced, even if it is small. Therefore, if we wish to
test the various cosmologies up to very high redshifts we need to choose
a specific cosmology and implement the calibration procedure quoted above.
In this paper we are interested in testing the Cardassian cosmology with the
current GRBs data after calibrated them with the luminosity-distance relation 
predicted by this model.

\indent This paper  is organized as follows: in Section\,\ref{cardassian} we 
present an overview of the Cardassian model. In Section\,\ref{hdofgrb}\, we 
describe the way the GRBs calibration procedure is implemented and demonstrate
that it is self-consistent by comparing our result for the $\Lambda$-CDM scenario
with the result obtained by Schaefer (2007). Then the construction of the Hubble 
diagram (HD) of those GRBs is made. In Section\,\ref{chi2analysis} we perform the 
best-fit analysis for several cases. In Section\, \ref{dinamical} we study other
observational aspects of the model like periods of both Cardassian and acelerate 
expansion and the age of universe. Finally, in Section\,\ref{conclusiones} we 
present our discussion and conclusions.

\section{\label{cardassian} Cardassian Cosmology}

The dynamics of the  universe is worked out through both the Friedmann's
equation and the evolution equations for a perfect, pressureless and
non-interacting fluid

\begin{eqnarray}
\label{fried1}
H^2 = \left(\frac{\dot{a}}{a}\right)^2 & = & \frac{8\pi G_N }{3}\rho_i + 
\frac{k}{a^2} \; , \\
\label{conen}
\dot{\rho_i}+3\,H(\rho_i +p_i)=0 &\Rightarrow & \rho_i = \rho_{i,0}
\left(\frac{a_0}{a}\right)^3 \; .
\end{eqnarray}

Here the subscript ``$i$'' refers to the components of the ideal fluid,
$a$  is the scale factor and the subscript ``$0$'' refers to present-day
values. Inspired (apparently) on extra dimensions physics, Fresse and Lewis
\cite{free,yu} proposed an explanation for the acceleration phase without
invoking any vacuum energy or cosmological constant, and for a flat universe
model as required by the CMB observations. Thus, they proposed a modified
Friedmann equation, Eq.(\ref{fried1}): $H^2=g(\rho_m)$, \, with $k=0$, where
$g(\rho_m)$ is a different function of the energy density and $\rho_m$
contains only radiation and matter (including baryon and cold dark matter).
The function $g(\rho_m)$ returns to the usual term during the early history
of the universe, but takes a different form which allows to explain the
occurrence of an accelerating expansion in the recent past of the universe,
namely at $z\sim {\cal O}$(1).

The modified Friedmann's equation reads

\be
 H^2 = Ag(\rho_{\m}) = A\rho_{\m}\left[1 + \left(\frac{\rho_{card}}{\rho_{m}}
\right)^{(1-n)} \right] \; ,
 \label{eqcard}
\ee

where $A=8\pi\,G_N/3$. Thus, for $z < z_{card}$ the second term dominates\footnote{It
defines the Cardassian era}. This fact, together with the observational data, allows
to determine the value of the parameter $n$. In that case, the first term can be 
neglected and thus one finds that the scale factor evolves in time following the law:

 \begin{equation}
 a\propto t^{\frac{2}{3n}} \; ,
 \end{equation}

so that the (accelerate) expansion is superluminal for $n<2/3$. The case of
$n=2/3$ produces a term in the Friedmann' equation  $H^2\propto a^{-2}$, which
looks similar to a curvature term. This feature turns the model attractive
because the matter alone is sufficient to provide a flat geometry. Because
of the extra term on the right-hand side of the modified Friedmann's equation,
the critical mass density necessary to have a flat universe can be modified so
that the total density of the universe (see Eqs.(\ref{conen},\ref{eqcard})) reads

\be
\rho_{\text{total}}(z) = \rho_{\text{c,old}}\{\omm^{\rm obs}(1+z)^3+\oxx\,f_{\textrm{x}}(z)\} \; .
\label{dtot}
\ee

\indent Here $\rho_{total}(z) = \rho_{\m}(z) + \rho_{\textrm{x}}(z)$. Besides, $\rho_{c,old}=
3H_0^2/(8\pi G_N)$ is  the usual critical density in the standard FLRW cosmology
\footnote{$\rho_{c,old} = 1.878 \mathrm{x} 10^{-29}~ h^2_0~\dt\,$, this value is
given by the WMAP three year observations}, and $\omm^{obs}$ is the observed matter
density of the universe \footnote{we take $\omm^{obs}=0,27$ as our fiducial value}.
The subscript ``x'' refers to any component of the universe  that provides an additional
term in the Einstein's equations. Generically, it is called \emph{dark energy}, but in
the Cardassian case it is an additional matter term. For the Cardassian model both terms 
in Eq.(\ref{dtot}) come from matter, namely

 \be
 \Omega_{\textrm{total}} = \omm + \oxx = 1 \; ,
 \ee

where $\Omega_i$ are the fractional density of each component. The observed matter
density fraction today is given by the ratio of the critical mass density of the
Cardassian universe,  $\rho_{c,card}=\rho_{m,0}$ to that of the standard universe,
$\rho_{c,old}$. From Eq.(\ref{eqcard}), in the present time we obtain directly:

 \be
 \omm^{\rm obs} = \frac{\rho_{\m,0}}{\rho_{\text{c,old}}} = \frac{1}{[1 +
(1+z_{\text{card}})^{3(1-n)}]} \; .
 \ee

 \indent Conversely, we can express \zc,  and $\rho_{\textrm{card}}$ in terms of $\omm$

 \begin{eqnarray}
 z_{\textrm{card}}&=&\left[ (\omm^{{\rm obs} -1}-1\right]^{1/3(1-n)}-1 \label{zcarda} \; ,\\
 \rho_{\text{card}}&=&\rho_{\text{c,old}}\,\omm^{\rm obs}\left[\omm^{{\rm obs}-1}-1 
\right]^{1/(1-n)}\; .
\label{rhocarda}
 \end{eqnarray}

 Finally, from Eq.(\ref{dtot}), the dimensionless dark energy density $f_{\textrm{x}}(z)$ is
given by

 \be
 f_{\text{x}}(z)=\frac{\rho_{\textrm{x}}(z)}{\rho_{\textrm{x}}(0)}=(1+z)^{3n} \; .
 \ee

If the dark energy density corresponds to a cosmological constant, one finds that $f_{\textrm{x}}
(z)=1$, or equivalently $n = 0$ and $\omm = 0.27$ for all redshift $z$. 

Besides, in cosmology the luminosity distance is defined as \cite{sw,coles}

\be
 d_L=\frac{a_0^2}{a}r_1,
 \label{dlum}
 \ee

where $r_1$ is the co-moving coordinate of the source. Using the expression for
the propagation of light \,$ds^2=0=dt^2-\frac{a^2\,dr^2}{1-kr^2}$, \,the modified
Friedmann's equation (\ref{eqcard}) and Eq.(\ref{dtot}) we can re-cast the
luminosity distance for a flat universe as

\begin{eqnarray}
 \nonumber
 d_L(z) = (1+z)\frac{c}{H_0} \int_0^z \frac{dz'}{E(z')} \; ,
 \label{dlum1}
\end{eqnarray}

where $E^2(z) = \omm^{\rm obs}(1+z)^3 + \oxx\, f_{\textrm{x}}(z)$.

\indent For an outlying source of apparent $m$ and absolute $M$ magnitudes,
distance estimates are made through the distance-modulus $m-M$, which relates
to the luminosity distance (here expressed in units of \mega\parsec) through:

 \be
 \mu(z) = m - M = 5 \log{d_L(z)} + 25 \; .
 \label{dmodulo}
 \ee

By plotting the value of the distance modulus, $\mu(z)$, computed from the 
estimated luminosity distance, $d_L(z)$), as a function of the redshift ($z$) 
one can construct the Hubble diagram of gamma-ray bursts after calibrated with 
the Cardassina model.

\begin{figure}[t!]
\begin{center}
\includegraphics[scale=0.375]{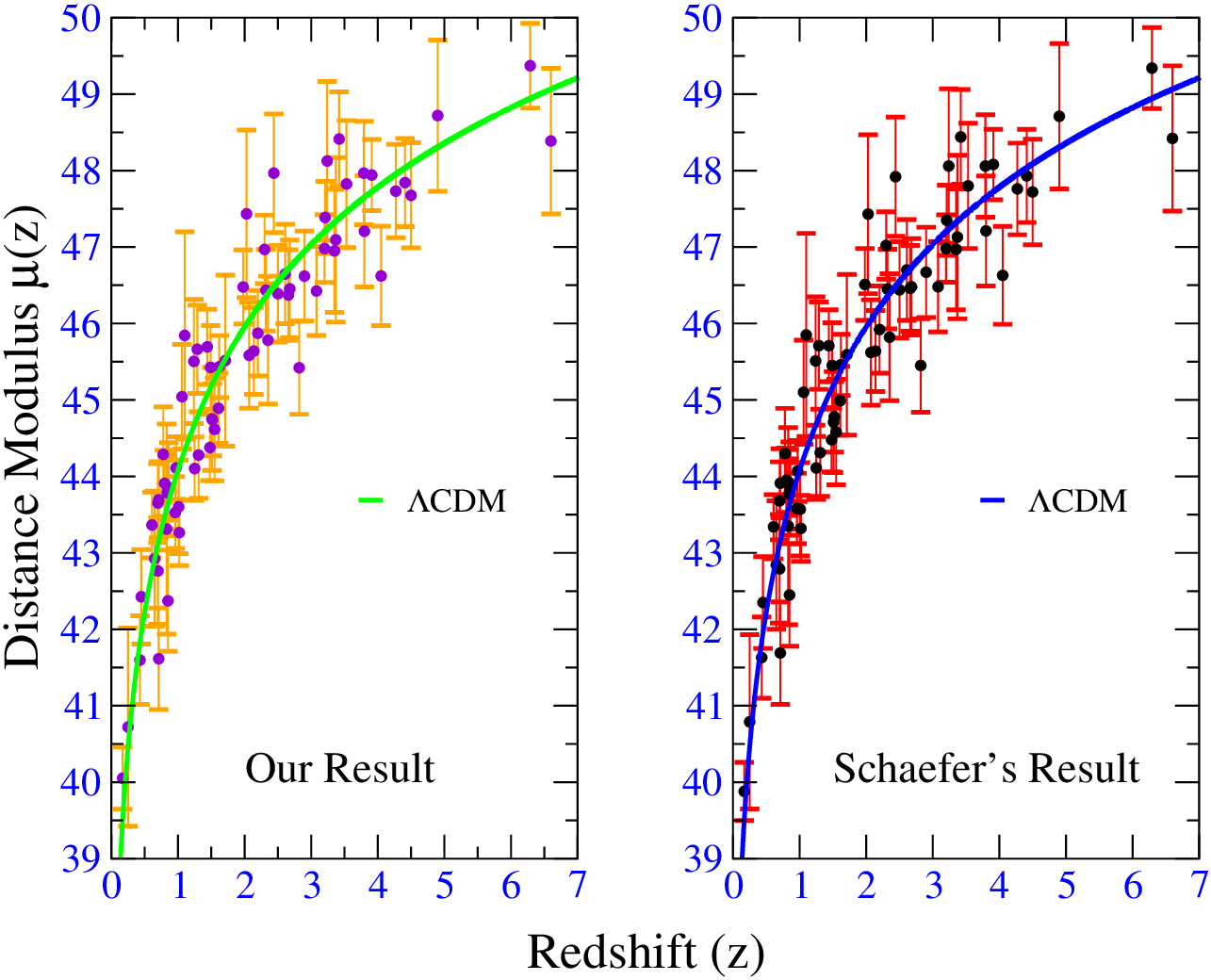}
\end{center}
\caption{\label{dhlcdm} Hubble Diagram of 69 GRBs calibrated with the
Concordance Cosmology (flat universe with $\omm = 0.27$ and $w=-1$), using
$H_0 = 72$ \phu .\, The left panel corresponds to our result (using the
standard OLS method \cite{isobe1}) and the right panel corresponds to
to Schaefer's procedure (Data taken from Column (8) of Table 6 in BS2007).}
\end{figure}

\section{\label{hdofgrb} Hubble diagram of gamma-ray bursts }

Gamma-ray bursts are the biggest explosions in the universe. The major
breakthrough in the understanding of GRBs came with the Beppo-SAX satellite
discovery of the X-ray afterglow of GRB970228 \cite{costa}. This allowed
the first precise determination of the a GRB event redshift, what definitely 
confirmed the long-standing suspicion that GRBs had cosmological origin. The 
huge power emitted during a GRB event makes GRBs detectable at z$\sim$ 20, or 
even higher, well-deep within the range of the epoch of reionization
\cite{schaefer2003a,ghirlanda2004a,dai,hooper}.

\indent As quoted above, the possibility of using GRBs as actual cosmological
probes stimulated the search for self-consistent methods of bringing GRBs into
the realm of cosmology. Empirical relations like the  Amati relation \cite{amati1},
the Ghirlanda relation \cite{ghirlanda2004a}, the Liang \& Zhang relation
\cite{liang2005} and the Firmani relation \cite{firmani2006} were introduced.
All of these relations take into account the
most relevant physical properties of GRBs: the peak energy,
denoted by \epk, which is the photon energy at which the
$\nu\,F_{\nu}$ spectrum is brightest; the jet openning angle,
denoted by $\theta_{\textrm{jet}}$, which is the rest-frame time
of the achromatic break in the light curve of an afterglow; the
time lag, denoted by $\tlag$, which measures the time offset
between high and low energy GRB photons arriving on Earth and the
variability, denoted by $V$, which is the measurement of the
``spikiness'' or ``smoothness'' of the GRB light curve.

Recently Schaefer in BS2007 presented a subset of luminosity vs. luminosity 
indicator relations that he used to calibrate GRBs with $\Lambda$-CDM. He 
also proposed an additional relation between minimum rise time and luminosity, 
where the minimum rise time, denoted by $\trt$, is taken to be the shortest 
time over which the light curve rises by half the peak flux of the pulse.

\indent For a GRB event to be placed on the HD it is necessary to know
its isotropic luminosity or its total collimation-corrected energy and redshift.
The first property can not be measured directly but rather it can be obtained
through the knowledge of either the bolometric peak flux, denoted by
\pbo\; ; or bolometric fluence; denoted by \sbo\, (BS2007). Therefore,
the isotropic luminosity is given by:

\be \label{liso}
 L = 4\pi\, d^2_{L}(z)P_{\text{bolo}} \; ,
\ee

and the total collimation-corrected energy reads:

\be
\label{egam}
E_{\gamma}=4\pi d^2_{L}(z)S_{\textrm{bolo}}F_{\text{beam}}(1+z)^{-1} \; ,
\ee

where $F_{\textrm{beam}}$ is the beaming factor ($1-\cos{\theta_{\textrm{jet}}}$).

The luminosity relations are power-law relations of either $L$ or
$E_{\gamma}$ as a function of $\tlag$, $V$,\, $E_{\text{peak}}$,\,
$\trt$. Both $L$ and $E_{\gamma}$ will be recalculated with
luminosity distances, and in the case of the Cardassian cosmology
with an appropriate choice of its cosmological parameters. 
\footnote{\label{cosmology-independent-analysis} The attentive 
reader should bear in mind that despite the assumption of a fiducial
cosmology to start with, after the calibration procedure one ends up
with almost a cosmology-independent result, i. e., {\it no circularity
problem}\cite{bs2007}.}

\begin{figure}[t!]
\begin{center}
\includegraphics[scale=0.375]{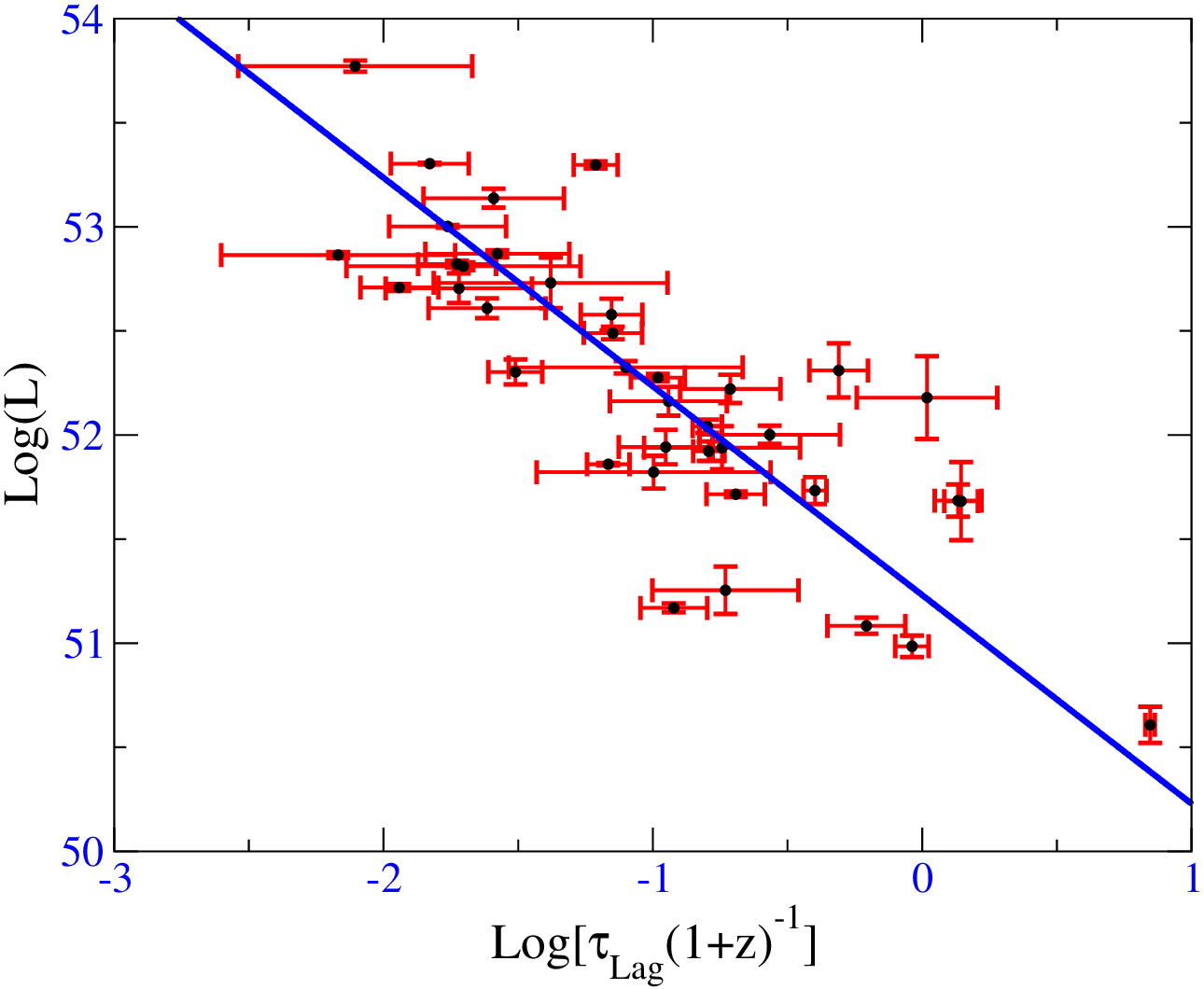}
\end{center}
\caption{\label{cali1} Time lag-luminosity relation. The time lags for 39 GRBs 
were corrected to the rest frame of each GRB event and plotted versus the isotropic 
luminosity with the best-fit (see Eq.(\ref{eqcal1})) line superposed on them. 
Notice that the slope in the figure is nearly the theoretical value predicted 
to be $-1$.}
\end{figure}

\indent Table-\ref{tablebs2007} below is the same as Table-4 of 
BS2007. Schaefer has compiled all the properties needed to calibrate 
the GRB events, which then produces the GRBs HD for any particular
cosmology. Column (1) gives the six-digit GRB identification number 
in the usual format of YYMMDD. Column (2) gives the redshift ($z$) 
of the GRB rounded to the nearest 0.01. Column (3) is  \pbo\, and 
its uncertainty in units of \upb.\, Column (4) gives \sbo\, and its 
uncertainty in units of \usb.\, Column (5) gives the beaming factor,
$F_{\text{beam}}$. Column (6) gives the time lag for each burst in
units of seconds in the Earth rest frame and its uncertainty.
Column (7) gives the variability, $V$, and its uncertainty. Column
(8) lists the observed \epk\, and its uncertainty, and finally
Column (9) gives the minimun rise time, \,$\trt$\,,  in seconds in
the Earth rest frame.

 \begingroup
 \squeezetable
 \begin{table*}
 \caption{\label{tablebs2007} Observables attained from Schaefer's 69 GRBs sample}
 \begin{ruledtabular}
 \begin{tabular}{c c c c c c c c c}
   &    & \pbo  & \sbo &                  &   $\tlag$      &   & \epk\footnotemark[1]  & $\trt$  \\
GRB& $z$& (\upb)&(\usb)&$F_{\text{beam}}$ &    (\s)        &$V$&      (\kev)         &  (\s)       \\
(1)&(2) &   (3) & (4)  & (5)              &   (6)          &(7)&       (8)           &  (9)         \\
 \hline
\\
970228\bd\bd&0.70 &7.3E-6 $\pm$ 4.3E-7 & \cd                 & \cd               &\cd           &0.0059$\pm$0.0008 &115$_{-38}^{+38}$      & 0.26 $\pm$ 0.04\\
970508\bd\bd&0.84 &3.3E-6 $\pm$ 3.3E-7 &8.09E-6 $\pm$ 8.1E-7 &0.0795$\pm$ 0.0204 &0.50$\pm$0.30 &0.0047$\pm$0.0009 &389$_{-[40]}^{+[40]}$  & 0.71 $\pm$ 0.06\\
970828\bd\bd&0.96 &1.0E-5 $\pm$ 1.1E-6 &1.23E-4 $\pm$ 1.2E-5 &0.0053$\pm$ 0.0014 &\cd           &0.0077$\pm$0.0007 &298$_{-[30]}^{+[30]}$  & 0.26 $\pm$ 0.07\\
971214\bd\bd&3.42 &7.5E-7 $\pm$ 2.4E-8 & \cd                 & \cd               &0.03$\pm$0.03 &0.0153$\pm$0.0006 &190$_{-[20]}^{+[20]}$  & 0.05 $\pm$ 0.02\\
980613\bd\bd&1.10 &3.0E-7 $\pm$ 8.3E-8 & \cd                 & \cd               &\cd           &\cd               &92 $_{-42  }^{+42  }$  & \cd            \\
980703\bd\bd&0.97 &1.2E-6 $\pm$ 3.6E-8 &2.83E-5 $\pm$ 2.9E-6 &0.0184$\pm$ 0.0027 &0.40$\pm$0.10 &0.0064$\pm$0.0003 &254$_{-[25]}^{+[25]}$  & 3.60 $\pm$ 0.5 \\
990123\bd\bd&1.61 &1.3E-5 $\pm$ 5.0E-7 &3.11E-4 $\pm$ 3.1E-5 &0.0024$\pm$ 0.0007 &0.16$\pm$0.03 &0.0175$\pm$0.0001 &604$_{-[60]}^{+[60]}$  & \cd            \\
990506\bd\bd&1.31 &1.1E-5 $\pm$ 1.5E-7 & \cd                 & \cd                 &0.04$\pm$0.02 &0.0131$\pm$0.0001 &283$_{-[30]}^{+[30]}$  & 0.17 $\pm$ 0.03\\
990510\bd\bd&1.62 &3.3E-6 $\pm$ 1.2E-7 &2.85E-5 $\pm$ 2.9E-6 &0.0021$\pm$ 0.0003 &0.03$\pm$0.01 &0.0100$\pm$0.0001 &126$_{-[10]}^{+[10]}$  & 0.14 $\pm$ 0.02\\
990705\bd\bd&0.84 &6.6E-6 $\pm$ 2.6E-7 &1.34E-4 $\pm$ 1.5E-5 &0.0035$\pm$ 0.0010 &\cd           &0.0210$\pm$0.0008 &189$_{-15  }^{+15  }$  & 0.05 $\pm$ 0.02\\
990712\bd\bd&0.43 &3.5E-6 $\pm$ 2.9E-7 &1.19E-5 $\pm$ 6.2E-7 &0.0136$\pm$ 0.0034 &\cd           &\cd               &65 $_{-10  }^{+10  }$  & \cd            \\
991208\bd\bd&0.71 &2.1E-5 $\pm$ 2.1E-6 & \cd                 & \cd               &\cd           &0.0037$\pm$0.0001 &190$_{-[20]}^{+[20]}$  & 0.32 $\pm$ 0.04\\
991216\bd\bd&1.02 &4.1E-5 $\pm$ 3.8E-7 &2.48E-4 $\pm$ 2.5E-5 &0.0030$\pm$ 0.0009 &0.03$\pm$0.01 &0.0130$\pm$0.0001 &318$_{-[30]}^{+[30]}$  & 0.08 $\pm$ 0.02\\
000131\bd\bd&4.50 &7.3E-7 $\pm$ 8.3E-8 & \cd                 & \cd               &\cd           &0.0053$\pm$0.0006 &163$_{-13  }^{+13  }$  & 0.12 $\pm$ 0.06\\
000210\bd\bd&0.85 &2.0E-5 $\pm$ 2.1E-6 & \cd                 & \cd               &\cd           &0.0041$\pm$0.0004 &408$_{-14  }^{+14  }$  & 0.38 $\pm$ 0.06\\
000911\bd\bd&1.06 &1.9E-5 $\pm$ 1.9E-6 & \cd                 & \cd               &\cd           &0.0235$\pm$0.0014 &986$_{-[100]}^{+[100]}$& 0.05 $\pm$ 0.02\\
000926\bd\bd&2.07 &2.9E-6 $\pm$ 2.9E-7 & \cd                 & \cd               &\cd           &0.0134$\pm$0.0013 &100$_{-7   }^{+7   }$  & 0.05 $\pm$ 0.03\\
010222\bd\bd&1.48 &2.3E-5 $\pm$ 7.2E-7 &2.45E-4 $\pm$ 9.1E-6 &0.0014$\pm$ 0.0001 &\cd           &0.0117$\pm$0.0003 &309$_{-12  }^{+12  }$  & 0.12 $\pm$ 0.03\\
010921\bd\bd&0.45 &1.8E-6 $\pm$ 1.6E-7 & \cd                   & \cd               &0.90$\pm$0.30 &0.0014$\pm$0.0015 &89 $_{-13.8}^{+21.8}$  & 3.90 $\pm$ 0.50\\
011211\bd\bd&2.14 &9.2E-8 $\pm$ 9.3E-9 &9.20E-6 $\pm$ 9.5E-7 &0.0044$\pm$ 0.0011 &\cd           &\cd               &59 $_{-8   }^{+8   }$  & \cd            \\
020124\bd\bd&3.20 &6.1E-7 $\pm$ 1.0E-7 &1.14E-5 $\pm$ 1.1E-6 &0.0039$\pm$ 0.0010 &0.08$\pm$0.05 &0.0131$\pm$0.0026 &87 $_{-12  }^{+18  }$  & 0.25 $\pm$ 0.05\\
020405\bd\bd&0.70 &7.4E-6 $\pm$ 3.1E-7 &1.10E-4 $\pm$ 2.1E-6 &0.0060$\pm$ 0.0020 &\cd           &0.0129$\pm$0.0008 &364$_{-90  }^{+90  }$  & 0.45 $\pm$ 0.08\\
020813\bd\bd&1.25 &3.8E-6 $\pm$ 2.6E-7 &1.59E-4 $\pm$ 2.9E-6 &0.0012$\pm$ 0.0003 &0.16$\pm$0.04 &0.0131$\pm$0.0003 &142$_{-13  }^{+14  }$  & 0.82 $\pm$ 0.10\\
020903\bd\bd&0.25 &3.4E-8 $\pm$ 8.8E-9 & \cd                 & \cd               &\cd           &\cd               &2.6$_{-0.8 }^{+1.4 }$  & \cd            \\
021004\bd\bd&2.32 &2.3E-7 $\pm$ 5.5E-8 &3.61E-6 $\pm$ 8.6E-7 &0.0109$\pm$ 0.0027 &0.60$\pm$0.40 &0.0038$\pm$0.0049 &80 $_{-23  }^{+53  }$  & 0.35 $\pm$ 0.15\\
021211\bd\bd&1.01 &2.3E-6 $\pm$ 1.7E-7 & \cd                 & \cd               &0.32$\pm$0.04 &\cd               &46 $_{-6   }^{+ 8  }$  & 0.33 $\pm$ 0.05\\
030115\bd\bd&2.50 &3.2E-7 $\pm$ 5.1E-8 & \cd                 & \cd               &0.40$\pm$0.20 &0.0061$\pm$0.0042 &83 $_{-22  }^{+53  }$  & 1.47 $\pm$ 0.50\\
030226\bd\bd&1.98 &2.6E-7 $\pm$ 4.7E-8 &8.33E-6 $\pm$ 9.8E-7 &0.0034$\pm$ 0.0008 &0.30$\pm$0.30 &0.0058$\pm$0.0047 &97 $_{-17  }^{+27  }$  & 0.70 $\pm$ 0.20\\
030323\bd\bd&3.37 &1.2E-7 $\pm$ 6.0E-8 & \cd                 & \cd               &\cd           &\cd               &44 $_{-26  }^{+90  }$  & 1.00 $\pm$ 0.50\\
030328\bd\bd&1.52 &1.6E-6 $\pm$ 1.1E-7 &6.14E-5 $\pm$ 2.4E-6 &0.0020$\pm$ 0.0005 &0.20$\pm$0.20 &0.0053$\pm$0.0007 &126$_{-13  }^{+14  }$  & \cd            \\
030329\bd\bd&0.17 &2.0E-5 $\pm$ 1.0E-6 &2.31E-4 $\pm$ 2.0E-6 &0.0049$\pm$ 0.0009 &0.14$\pm$0.04 &0.0097$\pm$0.0002 &68 $_{-2.2 }^{+2.3 }$  & 0.66 $\pm$ 0.08\\
030429\bd\bd&2.66 &2.0E-7 $\pm$ 5.4E-8 &1.13E-6 $\pm$ 1.9E-7 &0.0060$\pm$ 0.0029 &\cd           &0.0055$\pm$0.0057 &35 $_{-8   }^{+12  }$  & 0.90 $\pm$ 0.20\\
030528\bd\bd&0.78 &1.6E-7 $\pm$ 3.2E-8 & \cd                 & \cd               &12.5$\pm$0.50 &0.0022$\pm$0.0019 &32 $_{-5.0 }^{+4.7 }$  & 0.77 $\pm$ 0.20\\
040924\bd\bd&0.86 &2.6E-6 $\pm$ 2.8E-7 & \cd                 & \cd               &0.30$\pm$0.04 &\cd               &67 $_{-6   }^{+6   }$  & 0.17 $\pm$ 0.02\\
041006\bd\bd&0.71 &2.5E-6 $\pm$ 1.4E-7 &1.75E-5 $\pm$ 1.8E-6 &0.0012$\pm$ 0.0003 &\cd           &0.0077$\pm$0.0003 &63 $_{-13  }^{+13  }$  & 0.65 $\pm$ 0.16\\
050126\bd\bd&1.29 &1.1E-7 $\pm$ 1.3E-8 & \cd                 & \cd               &2.10$\pm$0.30 &0.0039$\pm$0.0015 &47 $_{-8   }^{+23  }$  & 3.90 $\pm$ 0.80\\
050318\bd\bd&1.44 &5.2E-7 $\pm$ 6.3E-8 &3.46E-6 $\pm$ 3.5E-7 &0.0020$\pm$ 0.0006 &\cd           &0.0071$\pm$0.0009 &47 $_{-8   }^{+15  }$  & 0.38 $\pm$ 0.05\\
050319\bd\bd&3.24 &2.3E-7 $\pm$ 3.6E-8 & \cd                 & \cd               &\cd           &0.0028$\pm$0.0022 &  \cd               & 0.19 $\pm$ 0.04\\
050401\bd\bd&2.90 &2.1E-6 $\pm$ 2.2E-7 & \cd                 & \cd               &0.10$\pm$0.06 &0.0135$\pm$0.0012 &118$_{-18  }^{+ 18 }$  & 0.03 $\pm$ 0.01\\
050406\bd\bd&2.44 &4.2E-8 $\pm$ 1.1E-8 & \cd                 & \cd               &0.64$\pm$0.40 &\cd               &25 $_{-13  }^{+ 35 }$  & 0.50 $\pm$ 0.30\\
050408\bd\bd&1.24 &1.1E-6 $\pm$ 2.1E-7 & \cd                 & \cd               &0.25$\pm$0.10 &\cd               &\cd                    & 0.25 $\pm$ 0.08\\
050416\bd\bd&0.65 &5.3E-7 $\pm$ 8.5E-8 & \cd                 & \cd               &\cd           &\cd               &15 $_{-2.7 }^{+ 2.3}$  & 0.51 $\pm$ 0.30\\
050502\bd\bd&3.79 &4.3E-7 $\pm$ 1.2E-7 & \cd                 & \cd               &0.20$\pm$0.20 &0.0221$\pm$0.0029 &93 $_{-35  }^{+ 55 }$  & 0.40 $\pm$ 0.20\\
050505\bd\bd&4.27 &3.2E-7 $\pm$ 5.4E-8 &6.20E-6 $\pm$ 8.5E-7 &0.0014$\pm$ 0.0007 &\cd           &0.0035$\pm$0.0019 &70 $_{-24  }^{+ 140}$  & 0.40 $\pm$ 0.15\\
050525\bd\bd&0.61 &5.2E-6 $\pm$ 7.2E-8 &2.59E-5 $\pm$ 1.3E-6 &0.0025$\pm$ 0.0010 &0.11$\pm$0.02 &0.0135$\pm$0.0003 &81 $_{-1.4 }^{+ 1.4}$  & 0.32 $\pm$ 0.03\\
050603\bd\bd&2.82 &9.7E-6 $\pm$ 6.0E-7 & \cd                 & \cd               &0.03$\pm$0.03 &0.0163$\pm$0.0015 &344$_{-52  }^{+ 52 }$  & 0.17 $\pm$ 0.02\\
050802\bd\bd&1.71 &5.0E-7 $\pm$ 7.3E-8 & \cd                 & \cd               &\cd           &0.0046$\pm$0.0053 &\cd                    & 0.80 $\pm$ 0.20\\
050820\bd\bd&2.61 &3.3E-7 $\pm$ 5.2E-8 & \cd                 & \cd               &0.70$\pm$0.30 &\cd               &246$_{-40  }^{+ 76 }$  & 2.00 $\pm$ 0.50\\
050824\bd\bd&0.83 &9.3E-8 $\pm$ 3.8E-8 & \cd                 & \cd               &\cd           &\cd               &\cd                    & 11.0 $\pm$ 2.00\\
050904\bd\bd&6.29 &2.5E-7 $\pm$ 3.5E-8 &2.00E-5 $\pm$ 2.0E-6 &0.0097$\pm$ 0.0024 &\cd           &0.0023$\pm$0.0026 &436$_{-90  }^{+ 200}$  & 0.60 $\pm$ 0.20\\
050908\bd\bd&3.35 &9.8E-8 $\pm$ 1.5E-8 & \cd                 & \cd               &\cd           &\cd               &41 $_{-5   }^{+ 9  }$  & 1.50 $\pm$ 0.30\\
050922\bd\bd&2.20 &2.0E-6 $\pm$ 7.3E-8 & \cd                 & \cd               &0.06$\pm$0.02 &0.0033$\pm$0.0006 &198$_{-22  }^{+ 38 }$  & 0.13 $\pm$ 0.02\\
051022\bd\bd&0.80 &1.1E-5 $\pm$ 8.7E-7 &3.40E-4 $\pm$ 1.2E-5 &0.0029$\pm$ 0.0001 &\cd           &0.0122$\pm$0.0004 &510$_{-20  }^{+ 22 }$  & 0.19 $\pm$ 0.04\\
051109\bd\bd&2.35 &7.8E-7 $\pm$ 9.7E-8 & \cd                 & \cd               &\cd           &\cd               &161$_{-35  }^{+ 130}$  & 1.30 $\pm$ 0.40\\
051111\bd\bd&1.55 &3.9E-7 $\pm$ 5.8E-8 & \cd                 & \cd               &1.02$\pm$0.10 &0.0024$\pm$0.0007 &\cd                    & 3.20 $\pm$ 1.00\\
060108\bd\bd&2.03 &1.1E-7 $\pm$ 1.1E-7 & \cd                 & \cd               &\cd           &0.0032$\pm$0.0058 &65 $_{-10  }^{+ 600}$  & 0.40 $\pm$ 0.20\\
\end{tabular}
\footnotetext[1]{We take this table  of BS2007. The uncertainties
given in square brackets are conservative estimates for the case
in which no error  bar is quoted in the original literature.}
   \end{ruledtabular}
 \end{table*}
 \endgroup

 \begingroup
 \squeezetable
 \begin{table*}[t!]
  \contcaption{Continued ...}
 \begin{ruledtabular}
 \begin{tabular}{c c c c c c c c c}
  &     & $P_{bolo}$&$S_{bolo}$&            &$\tlag$     &   &$E_{peak}$\footnotemark[1]& $\trt$  \\
GRB& $z$& (\upb)    &(\usb)    & $F_{beam}$ &   (\s)     &$V$&    (\kev)                 &  (\s)    \\
(1)& (2)&   (3)     & (4)      & (5)        &    (6)     &(7)&    (8)                    &  (9)      \\
 \hline
 \\
060115\bd\bd&3.53 &1.3E-7 $\pm$ 1.6E-8 & \cd                 & \cd               &\cd           &\cd               &62 $_{-6   }^{+ 19 }$  & 0.40 $\pm$ 0.20\\
060116\bd\bd&6.60 &2.0E-7 $\pm$ 1.1E-7 & \cd                 & \cd               &\cd           &\cd               &139$_{-36  }^{+ 400}$  & 1.30 $\pm$ 0.50\\
060124\bd\bd&2.30 &1.1E-6 $\pm$ 1.2E-7 &3.37E-5 $\pm$ 3.4E-6 &0.0021$\pm$ 0.0002 &0.08$\pm$0.04 &0.0140$\pm$0.0020 &237$_{-51  }^{+ 76 }$  & 0.30 $\pm$ 0.10\\
060206\bd\bd&4.05 &4.4E-7 $\pm$ 1.9E-8 & \cd                 & \cd                 &0.10$\pm$0.10 &0.0025$\pm$0.0016 &75 $_{-12  }^{+ 12 }$  & 1.25 $\pm$ 0.25\\
060210\bd\bd&3.91 &5.5E-7 $\pm$ 2.2E-8 &1.94E-5 $\pm$ 1.2E-6 &0.0005$\pm$ 0.0001 &0.13$\pm$0.08 &0.0019$\pm$0.0004 &149$_{-35  }^{+ 400}$  & 0.50 $\pm$ 0.20\\
060223\bd\bd&4.41 &2.1E-7 $\pm$ 3.7E-8 & \cd                 & \cd               &0.38$\pm$0.10 &0.0075$\pm$0.0033 &71 $_{-10  }^{+ 100}$  & 0.50 $\pm$ 0.10\\
060418\bd\bd&1.49 &1.5E-6 $\pm$ 5.9E-8 & \cd                 & \cd               &0.26$\pm$0.06 &0.0070$\pm$0.0005 &230$_{-[20]}^{+ [20]}$ & 0.32 $\pm$ 0.08\\
060502\bd\bd&1.51 &3.7E-7 $\pm$ 1.6E-7 & \cd                 & \cd               &3.50$\pm$0.50 &0.0010$\pm$0.0017 &156$_{-33  }^{+ 400 }$ & 3.10 $\pm$ 0.30\\
060510\bd\bd&4.90 &1.0E-7 $\pm$ 1.7E-8 & \cd                 & \cd     &\cd    &0.0028$\pm$0.0019 &95 $_{-[30]}^{+ [60]}$ & \cd            \\
060526\bd\bd&3.21 &2.4E-7 $\pm$ 3.3E-8 &1.17E-6 $\pm$ 1.7E-7 &0.0034$\pm$ 0.0014 &0.13$\pm$0.03 &0.0112$\pm$0.0039 &25 $_{-[5] }^{+ [5] }$ & 0.20 $\pm$ 0.05\\
060604\bd\bd&2.68 &9.0E-8 $\pm$ 1.6E-8 & \cd                 & \cd               &5.00$\pm$1.00 &\cd               &40 $_{-[5] }^{+ [5] }$ & 0.60 $\pm$ 0.20\\
060605\bd\bd&3.80 &1.2E-7 $\pm$ 5.5E-8 & \cd                 & \cd               &5.00$\pm$3.00 &\cd               &169$_{-[30]}^{+ [200]}$& 2.00 $\pm$ 0.50\\
060607\bd\bd&3.08 &2.7E-7 $\pm$ 8.1E-8 & \cd                 & \cd               &2.00$\pm$0.50 &0.0059$\pm$0.0014 &120$_{-17}^{+ 190}   $ & 2.00 $\pm$ 0.20\\
\end{tabular}
\end{ruledtabular}
\footnotetext[1]{We take this table from BS2007. The uncertainties
given in square brackets are conservative estimates for the case
in which no error bar is quoted in the original literature.}
\end{table*}
\endgroup


\subsection{Luminosity Relations and Calibration Procedure}

The relationship between a measurable observable of the light curve
(luminosity indicator) with the GRB luminosity is given by the relation 
luminosity in the form of a power-law, i. e., $L = B_{\text{lag}} 
\tlag^{a_{\text{lag}}}$,\, $L=B_{\text{v}}V^{a_{\text{v}}}$,\,$L = 
B_{\text{peak}}\epkk^{a_{\text{peak}}}$,\,$E_{\gamma} = B_{\gamma, \text{peak}}\epkk^{a_{\gamma,\text{peak}}}$ and $L=B_{\text{\tiny RT}}
\trt^{a_{\text{\tiny RT}}}$. The observed (on Earth) luminosity indicators 
will have different values from those that would be abscribed in the rest 
frame of the GRB event. That is, the light curves and spectra seen by the 
Earth-orbiting satellites suffer time dilation and redshifting. Therefore, 
the physical connection between the indicators and the luminosity in the 
GRB rest frame must take into account the observed indicators and
correct them to the rest frame of the GRB. For the temporal indicators,
the observed quantities must be divided by $1+z$ to correct the
time dilation. The observed $V$-value  must be multiplied by
$1+z$ because it varies inversely with time, and the observed
$\epkk$ must be multiplied by $1+z$ to correct the redshift
dilation of the spectrum. We have also used the same values employed
by BS2007 for the luminosity indicators to minimize correlations
between the normalization constant and the exponent during the
fitting, i. e., for the temporal luminosity we use 0.1\second, for
the variability 0.02, and for the energy indicator 300\,\kev. Notice 
that these values are only valid for this data. Therefore, for another
set of GRBs data we can not use them.

\begin{figure}[t!]
\begin{center}
\includegraphics[scale=0.375]{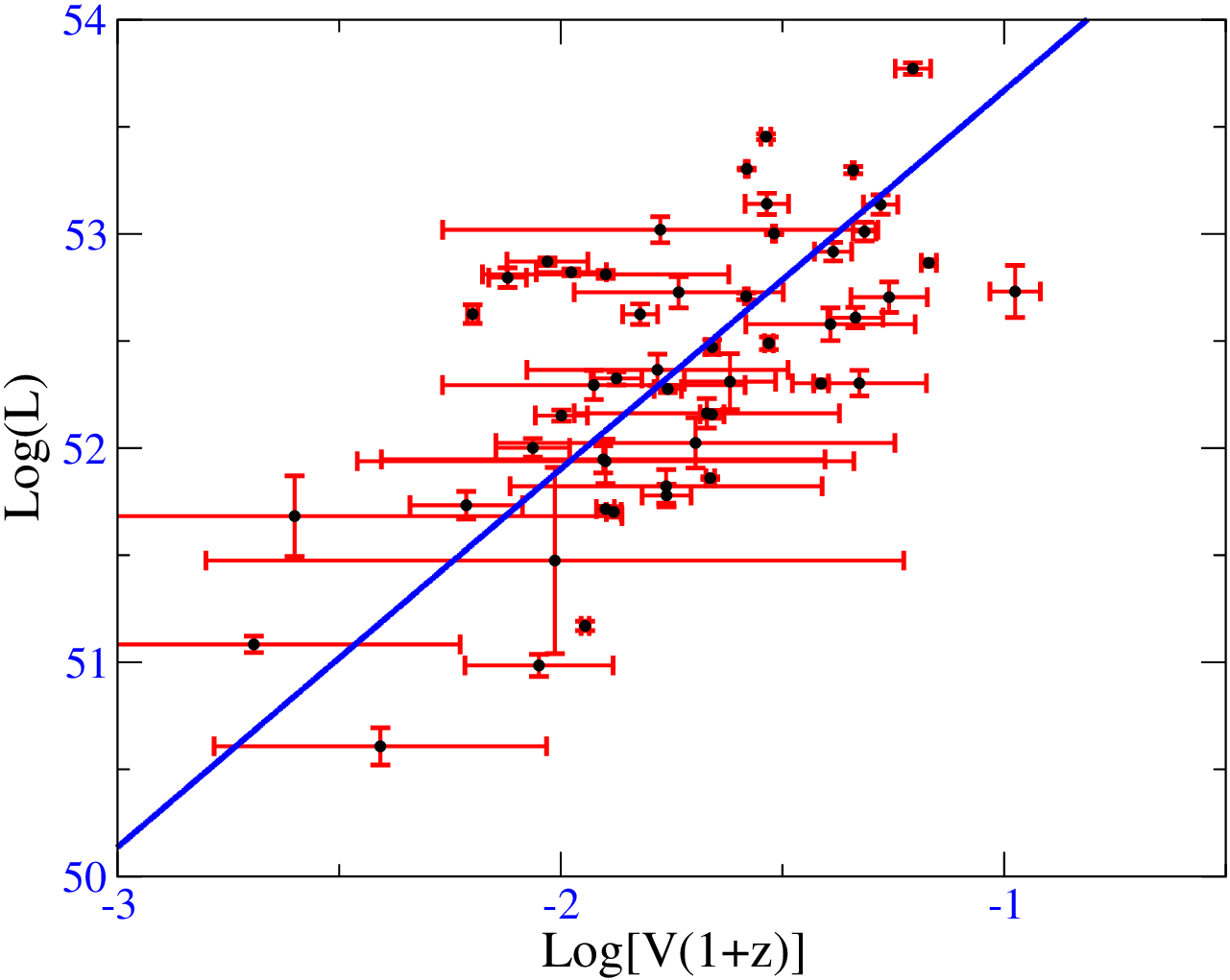}
\end{center}
\caption{\label{cali2} Variability-luminosity relation.
The variability for 51 GRBs have been corrected to rest
frame of each GRB event and plotted vs. the isotropic luminosity
with the best-fit line (see Eq.\ref{eqcal2}). This
relation was verified by Schaefer\cite{schaefer2001}.}
\end{figure}

\indent To explain the calibration procedure in general, we
denote the five luminosity relations by $R_i=B_i\,Q_i^{a_i}$
and we take their logarithms to express them as a linear relation
of the form

 \begin{displaymath}
 \log{R_i}=\log{B_i}+a_i\,\log{Q_i} \Rightarrow  y_i= b+a\,x_i \; .
\end{displaymath}

\indent In order to use the linear regression method to determine
the best-fit parameters we  also take into account that in those
luminosity relations these variables are independent. This feature
allows to use the OLS Bisector method \cite{isobe1}. Besides, because
the scatter of the data is consistent with a Gaussian distribution
we can use the error propagation law: \emph{The optimum $\bar{f}$ for
the quantity of interest $f(\alpha_i)\, i=1,2,\ldots n$ calculated
from the medians $\bar{\alpha_i}$ and their standard deviation
$\sigma_{\bar{f}}$ is given by:}\cite{error-propagation}

\be
\sigma_{\bar{f}}=\left[\sum_i^n\left (\frac{\partial f}{\partial
\alpha_i}\right)^2\sigma^2_{\bar{\alpha}_i} \right]^{1/2} \; .
\label{lpe}
\ee

\indent Then, for the luminosity indicators, applying this law, we
obtain the standard deviation associated to the $x$-variable

\begin{eqnarray}
 \nonumber
\sigma_{x} = \left(\frac{d\,x(Q_i)}{d\,Q_i}\right) \sigma_{\text{\tiny Q}_i} =
\frac{1}{\ln{10}} \frac{\sigma_{\text{\tiny Q}_i}}{Q_i} \; .
\label{sdx}
\end{eqnarray}

Similarly, the standard deviation associated with the $y$-variable
is function of $P_{\text{bolo}}$ when we use the burst luminosity
Eq.(\ref{liso}), or is function of \sbo \; and $F_{\text{beam}}$ if
we use the total collimation-corrected energy  Eq.(\ref{egam}).

In the first case

 \be
 \sigma_{y}= \frac{1}{\ln{10}}\frac{\sigma_{\text{\tiny 
P}_{\text{bolo}}}}{P_{\text{bolo}}}\; .
 \label{sdliso}
 \ee

For the second case, corresponding to $\log{E_{\gamma}}$, the 
standard deviation is given by

\be
\sigma_y = \frac{1}{\ln{10}}\sqrt{\left(\frac{\sigma_{\text{\tiny 
S}_{\text{bolo}}}}{\text{S}_{\text{bolo}}}\right)^2
+ \left(\frac{\sigma_{\text{\tiny F}_{\text{beam}}}}{F_{\text{beam}}}
\right)^2} \; .
\label{sdegama}
\ee

\indent Then, the calibration will essentially be a linear fit on a
log-log plot of the luminosity indicator $Q_i$  versus the burst
luminosity $R_i$, i. e., we fit the line $ y_i= b+a\,x_i$, and find
the corrected value of the luminosity using the subroutine
\emph{Sixlin.f} \cite{isobe2}\, upon computing $\bar{y}_i(a,b,x_i)$.
Therefore, their associate error or standard deviation, using the
Eq.(\ref{lpe}), is given as

\be
\sigma^2_{\bar{y}_i} = \sigma^2_a + (\sigma_b\,x_i)^2 + (b\sigma_{x_i})^2 \; ,
 \label{sdy}
\ee

\noindent where $\sigma_{x_i}$ is given by Eq.(\ref{sdx}).\, As the
uncertainties in both luminosities and their indicators are small
than the observed scatter about the best-fit line, we must
introduce an additional source of intrinsic scatter in the last
expression, denoted by $\sigma_{\text{\tiny sys}}$. This value can
be estimated by finding the value such that a $\chi^{2}$ fit of
the luminosity calibration produces a reduced $\chi^{2}$ of unity

\be
\chi^2 = \frac{1}{N}\sum_i^N \frac{(\bar{y}_i-y_i)^2}{\sigma^2_{
\bar{y}_i} + \sigma^2_{\text{\tiny sys}}}\approx 1 \; .
\ee

Here $\sigma^2_{\bar{y}_i}$ is given by Eq.(\ref{sdy}) and then we
re-cast the standard deviation for the correct luminosity as:

\be
\sigma^2_{\bar{y}_i} = \sigma^2_a +
(\sigma_b\,x_i)^2 + (b\sigma_{x_i})^2 + \sigma^2_{\text{\tiny sys}} \; .
\label{sdy1}
\ee

\indent With the purpose of verifying our procedure we used the standard
OLS method \cite{isobe1} in a barely different form to that one
used by Schaefer (see pag.26 of BS2007), and then we calibrated the GRBs
for the Concordance Cosmology (flat universe with $\omm = 0.27$
and $w=-1$). We choose $H_0 = 72$~\phu\, from the HST Key project\cite{hst}.
In Fig.\ref{dhlcdm} we show our results and we compare them with
the HD obtained by Schaefer (2007). One can notice that the difference
between both plots is very small, what confirms that our method is
self-consistent.

\begin{figure}[t!]
\begin{center}
\includegraphics[scale=0.375]{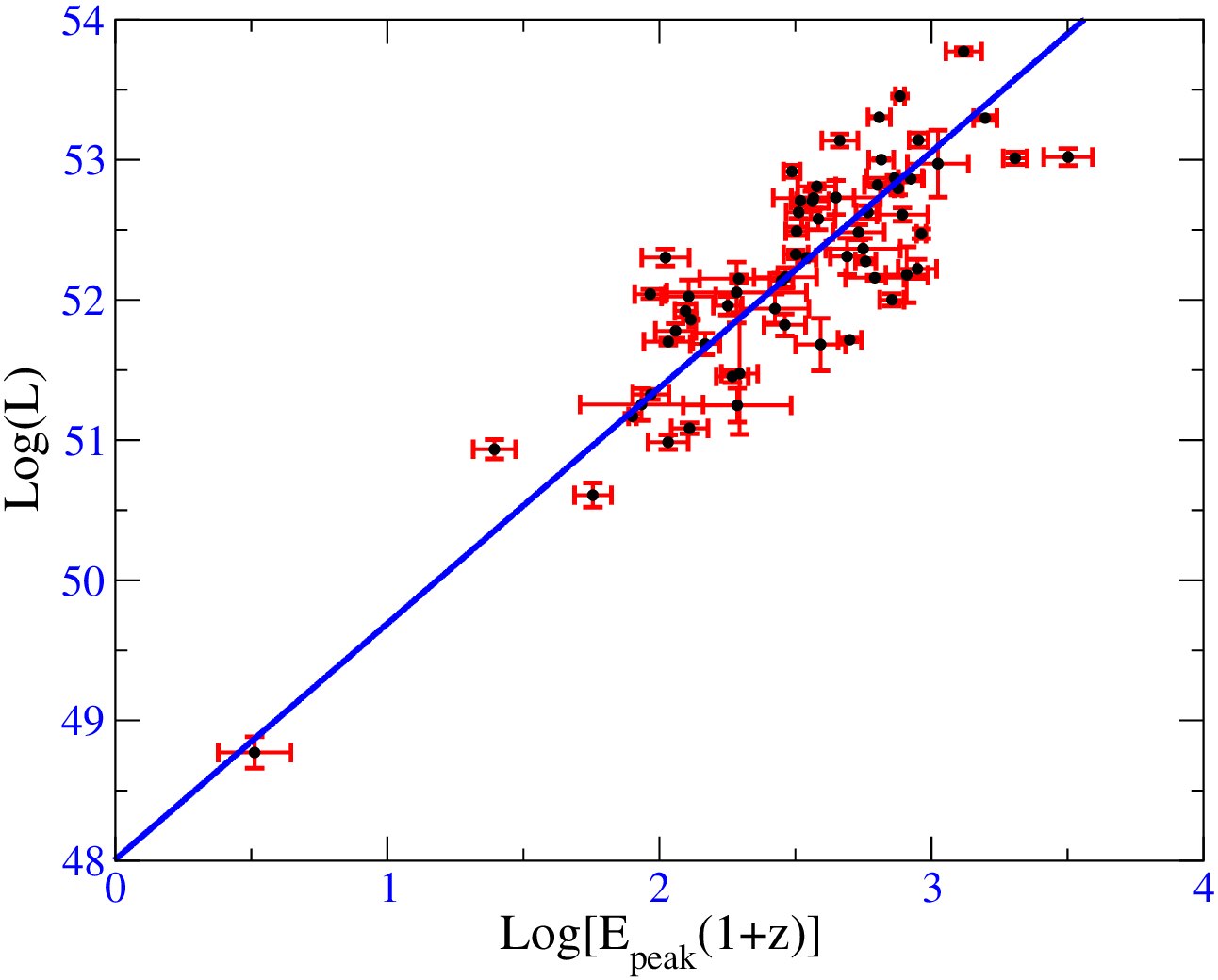}
\end{center}
\caption{\label{cali3} \epk-luminosity relation.
The \epk values for 64 GRBs have been corrected
to the rest frame of each GRB event and plotted
versus the isotropic luminosity with the best-fit line,
(see Eq.\ref{eqcal3}). This relation was proposed
and verified by Schaefer\cite{schaefer2003b,bs2007}.}
\end{figure}

\indent In the next subsections we present the results found with 
the calibration procedure of the five luminosity relations. All of
them are based on an assumed Cardassian model using Eq.(\ref{dlum1})
with $\omm=0.27,\,n=0.2$ and $H_0$ = 72\; \phu \ref{cosmology-independent-analysis}.

\subsubsection{\label{primera} Time lag versus Luminosity}

The time lag, $\tau_{\text{lag}}$, was identified as a luminosity indicator
by Norris et al. \cite{norris2000} who proposed a power law relation between
the time lag and the isotropic luminosity. This relation is a consequence
of the empirical/theoretical Liang-Kargatis relation \cite{liang1996}
and was verified by Schaefer \cite{schaefer2001}.
The calibration data are plotted in Fig.\ref{cali1} as
$\log{\tau_{\text{lag}}/(1+z)}$ versus $L$. The best-fit linear
regression line is also plotted. The equation for this calibration
line reads:

\be
\log{L} = 52.23 -1.00 \log\left[{\frac{\tlag}{(1+z) 0.1\second}}\right] \; .
\label{eqcal1}
\ee

Notice that the slope value satisfactorily agrees with the theoretical value of -1.
The 1-$\sigma$ uncertainties in the intercept ($a=52.23$) and slopes ($b=-1.00$)
are $\sigma_{a}=0.07$ and $\sigma_{b}=0.09$, and the uncertainty on the value of
$\log{L}$  using Eq.(\ref{sdy1}) (the $\sigma^2_{\text{lag,sys}}$-value found
is 0.36) is given by

 \begin{eqnarray}
 \sigma^{2}_{\log{L}}& = & \sigma^2_a+\left\{\sigma_b\left[\frac{\tlag}{(1+z)0.1s}
\right]\right\}^2 \; , \nonumber \\
 \nonumber \\
 & +  & \left(\frac{0.4343\,b\,\sigma_{\text{lag}}}{\tlag}\right)^2 + 
\sigma^2_{\text{lag,sys}} \; .
\label{sdcal1}
 \end{eqnarray}

\subsubsection{\label{segunda} Variability versus Luminosity}

The variability ($V$) of a  GRB event was identified as indicator luminosity by 
Fenimore \& Ramirez-Ruiz \cite{fenimore2000}. Subsequently, Reitchart \cite{reichart2001}
proposed a new relation between variability and isotropic luminosity, which is
similar to the time lag-luminosity. Its origin is based in the physics of the
relativistic shocked jets\cite{meszaros2002,kobayashi2002}. The calibration plot
for the $V-L$ relation is given in Fig.\ref{cali2} along with the best-fit line.
This best-fit can be represented by the equation

\be
\log{L} = 52.43 + 1.77\log\left[{\frac{V\,(1+z)}{0.02}}\right]\; .
\label{eqcal2}
\ee

The 1-$\sigma$ uncertainties in the intercept and slope are $\sigma_{a}=0.07$
and $\sigma_{b}=0.19$, the $\sigma_{V,\text{sys}}$ value found was 0.47 and
the uncertainty in the log of the luminosity, using Eq.(\ref{sdy1}) is

\be
 \sigma^{2}_{\log{L}} = \sigma^2_a + \left\{\sigma_b \left[\frac{V\,(1+z)}{0.02}
\right]\right\}^2 + \left(\frac{0.4343 \,b\, \sigma_{\text{v}}}{V}\right)^2 +
\sigma^2_{\text{v,sys}}\; .
\label{sdcal2}
\ee

\begin{figure}[t!]
\begin{center}
\includegraphics[scale=0.375]{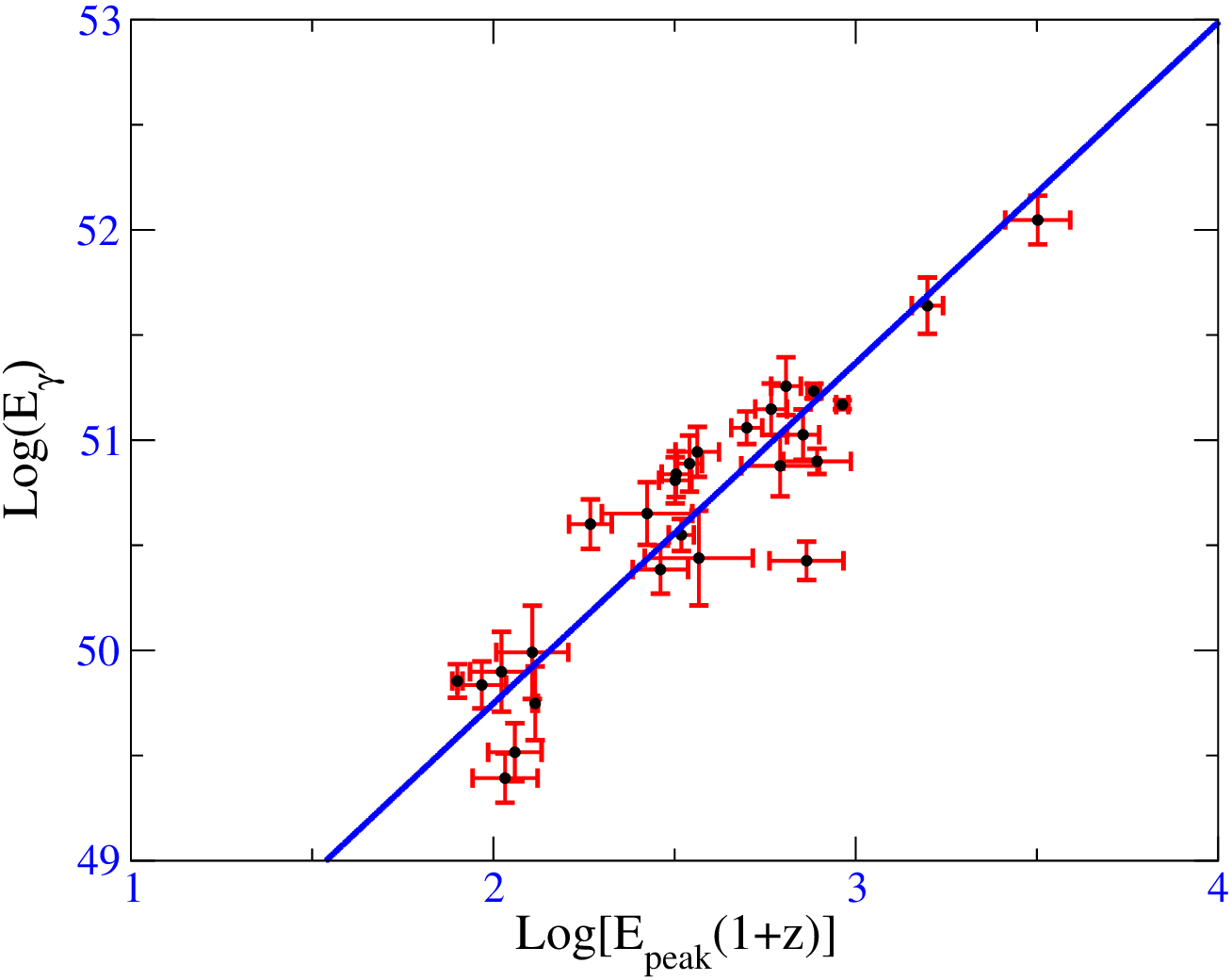}
\end{center}
\caption{\label{cali4} \epk-E$_{\gamma}$\, relation, or Ghirlanda
relation. The \epk \, values for 27 GRBs have been corrected to the
rest frame of the GRB event and plotted versus the total burst energy 
in the $\gamma$-rays with the best-fit line (see Eq.\ref{eqcal4}). This
relation allows an empirical correction to the determined luminosity
distance for each GRB event, in the same way as the light-curve shape
correction for SNIa. }
\end{figure}

\subsubsection{\label{tercera} \epk \, versus Luminosity}

This luminosity relation was proposed by Schaefer\cite{schaefer2003b} and
is related to the instantaneus physics at the time of the peak. It was
also verified in BS2007. The calibration plot for the $\epkk-L$ relation
is given in Fig.\ref{cali3} along with the best-fit line. This best-fit
line can be represented by the equation

\be
\log{L} = 52.18 + 1.68 \log\left[{\frac{\epkk\,(1+z)}{300\,\kev}}\right] \; .
\label{eqcal3}
\ee

The 1-$\sigma$ uncertainties in the intercept and slope are
$\sigma_{a}=0.05$ and $\sigma_{b}=0.10$, and the
$\sigma_{\epkk,\text{sys}}$ value is 0.4. The uncertainty
in the log of the luminosity using Eq.(\ref{sdy1}) is

\begin{eqnarray}
\sigma^{2}_{\log{L}}& =& \sigma^2_a+\left\{\sigma_b
\left[\frac{\epkk\,(1+z)}{300\,\kev}\right]\right\}^2  \nonumber \\
 \nonumber \\
 & & +\left(\frac{0.4343\,b\,\sigma_{\epkk}}{\epkk}\right)^2
+ \sigma^2_{\epkk,\text{sys}} \; .
 \label{sdcal3}
\end{eqnarray}

\subsubsection{\label{cuarta}\epk \,versus $E_{\gamma}$}

Ghirlanda et al. \cite{ghirlanda2004a} found that for GRBs the total
energy emitted in $\gamma$-rays (E$_{\gamma}$) after a proper
collimated correction, correlates tightly with the peak  energy
\epk\; (in the $\nu$-F$_{\nu}$ spectrum). Thus the isotropically
equivalent burst energy could be determined with sufficient accuracy
to be used in a practically fashion for cosmological studies.
The physics of this relation is explained within the standard
jet model\cite{eichler2004,yamazaki2004,rees2005,levinson2005}.
The calibration plot for the \epk-E$_{\gamma}$ relation is given in
Fig.\ref{cali4} along with the best-fit line. This best-fit can be
represented with the equation

\be
\log{E_{\gamma}} = 50.52 + 1.68 \log\left[{\frac{E_{peak}\,(1+z)}{300\,\kev}}\right]\; .
\label{eqcal4}
\ee

The 1-$\sigma$ uncertainties in the intercept and slope are
$\sigma_{a}=0.05$ and $\sigma_{b}=0.10$, and $\sigma_{\epkk,\text{sys}}
=0.21$. The uncertainty in the log of the luminosity is obtained using
the Eq.(\ref{sdy1}) is

\begin{eqnarray}
\sigma^{2}_{\log{E_{\gamma}}} & = &\sigma^2_a+\left\{\sigma_b\left[\frac{\epkk\,(1+z)}{300\,\kev}\right]\right\}^2
\nonumber\\
 \nonumber\\
& + & \left(\frac{0.4343\,b\,\sigma_{\epkk}}{\epkk}\right)^2+\sigma^2_{\epkk,\text{sys}} \; .
 \label{sdcal4}
\end{eqnarray}

\subsubsection{\label{quinta} Rise Time versus Luminosity}

\begin{figure}[t!]
\begin{center}
\includegraphics[scale=0.375]{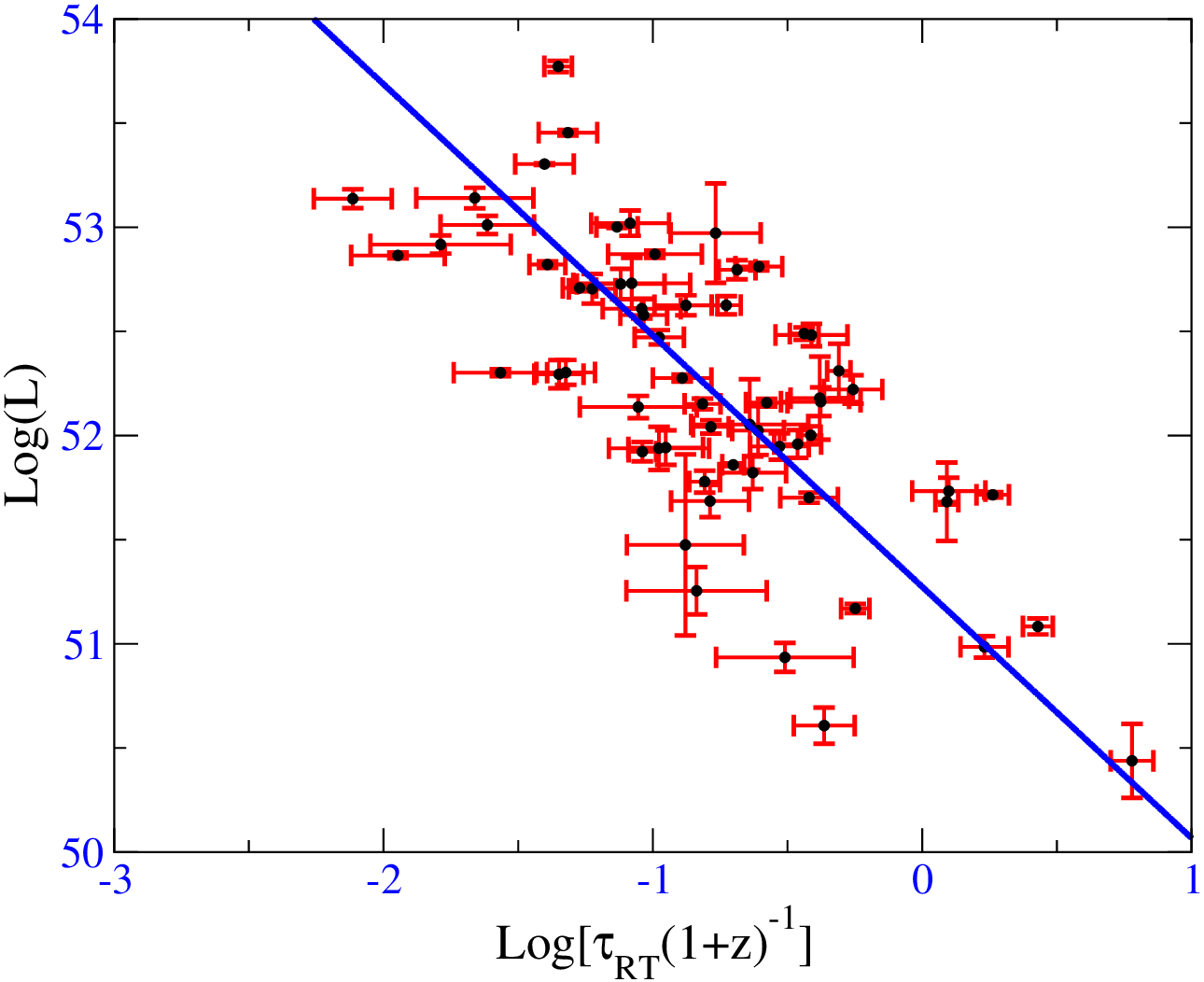}
\end{center}
\caption{\label{cali5} Minimum rise time-luminosity relation.
The rise time for 62 GRBs have been corrected to the rest frame
ofthe GRB event and plotted versus the isotropic luminosity with 
the best-fit line (see Eq.\ref{eqcal5}). This relation was proposed
and confirmed for Schaefer in BS2007.}
\end{figure}

In BS2007, in an effort to understand the physical origin of the
variability, Schaefer calculated the variability for a wide
range of simulate light curves constructed from individual pulses.
He found that the most important determinant of the V-value was the
rise time in the light curves, and this rise time can be connected
with the physics of the shocked jet.
The calibration plot for the $\tau_{\text{\tiny RT}}-L$ relation
is given in Fig.\ref{cali5} along with the best-fit line. This
best-fit can be represented with the equation

\be
\log{L} = 52.48 - 1.21 \log \left[{\frac{\tau_{\text{\tiny RT}}}{(1+z)\,0.1s}}\right]\; .
\label{eqcal5}
\ee

The 1-$\sigma$ uncertainties in the intercept and slope are $\sigma_{a}=0.07$ and
$\sigma_{b}=0.11$, and the value of $\sigma_{RT,sys}=0.47$. The uncertainty in the
log of the luminosity is given by Eq.(\ref{sdy1})

\begin{eqnarray}
 \sigma^{2}_{\log{L}}& = & \sigma^2_a+\left\{\sigma_b\left[\frac{\tau_{\text{
\tiny RT}}}{(1+z)\,0.1s} \right]\right\}^2  \nonumber\\
 \nonumber \\
 \nonumber\\
 & + &\left(\frac{0.4343\,b\, \sigma_{\tau_{\text{\tiny RT}}}} {\tau_{
\text{\tiny RT}}}\right)^2 + \sigma^2_{\text{\tiny RT,sys}} \; .
 \label{sdcal5}
\end{eqnarray}

\indent In Table-\ref{tabrelac} we collect our results. Notice
that they are very similar to the results obtained in BS2007 for
the Concordance Cosmology ($\omm=0.27$ and $w=-1$) and for the dark
energy parametrization ($w_0=-1.31,w^{'}=1.48$). Indeed, the set of
$\chi^2$ for both methods are as follows: $\chi^2{\rm \Lambda-CDM/Schafer}
= 71.61248$, $\chi^2{\rm \Lambda-CDM/Schafer}_{\rm dof} = 1.03786$, 
$\chi^2{\rm \Lambda-CDM/Ours} = 72.16559$, $\chi^2{\rm \Lambda-CDM/
Ours}_{\rm dof} = 1.04587$, $\chi^2{\rm Card} = 72.034$,
$\chi^2{\rm Card}_{\rm dof} = 1.04379$. This confirms one
more time that the calibration procedure depends weekly of the
input cosmology.

\begin{table}[t!]
 \caption{\label{tabrelac}Calibration Results}
  \begin{ruledtabular}
 \begin{tabular}{l c c c  c c }
 Luminosity Relation        & $a$   & $\sigma_a$  & $b$   & $\sigma_b$ & $\sigma_{\text{\tiny sys}}$\\
                            &       &             &       &            &                             \\
 \hline
 $\tau_{\text{lag}}-L$      & 52.23 &  0.07       & -1.00 &  0.09  &  0.36  \\
 $V-L$                      & 52.43 &  0.07       &  1.77 &  0.19  &  0.47  \\
 \epk-$L$                   & 52.18 &  0.05       &  1.68 &  0.10  &  0.40  \\
 \epk-E$_{\gamma}$          & 50.52 &  0.05       &  1.62 &  0.10  &  0.21  \\
 $\tau_{\text{\tiny RT}}-L$ & 52.48 &  0.07       & -1.21 &  0.11  &  0.47  \\
\end{tabular}
 \end{ruledtabular}
\end{table}

\subsubsection{Combining the Derived Distance Moduli}

After performing the calibration procedure for every $L$, or
$E_{\gamma}$, we can also derive a luminosity distance, or
equivalently their logarithms, and estimate their associate
distance moduli using Eqs.(\ref{dmodulo},\ref{liso},\ref{egam})
obtaining:

\begin{eqnarray}
\mu(z) & = & \frac{5}{2} \log{(L)} - \frac{5}{2} \log{(4\pi\,P_{\text{bolo}})} + 25 \; ,\\
 \nonumber \\
\mu(z) & =&\frac{5}{2}\log{(E_{\gamma})}-\frac{5}{2}\log{\left(\frac{4\pi\,S_{\text{bolo}}
\,F_{\text{beam}}}{1+z}\right)}\; .
\end{eqnarray}

The propagate uncertainties will depend on whether we calculate
the distance modulus using $L$ or $E_{\gamma}$. Therefore, using
Eq.(\ref{lpe}), one can easily obtain

\begin{equation}
 \sigma_{\mu} = 2.5 \, \left[\sg^2_{\bar{y}} + \sg^2_{y}\right]^{1/2}\; ,
\label{sdmu}
\end{equation}

\noindent where $\sg_{y}$ takes either the form of Eq.(\ref{sdliso}) if we
calculate $L$, or the form of the Eq.(\ref{sdegama}) if we
calculate $E_{\gamma}$, and $\sg_{\bar{y}}$ is given by
Eq.(\ref{sdy1}) applied to the respective luminosity relation.

\indent With five luminosity indicators, each burst will have up
to five estimated distance moduli an their 1-$\sigma$ uncertainties.
As in BS2007, we label as $\mu_1\pm
\sigma_{\mu_1}$,\, $\mu_2\pm \sigma_{\mu_2}$,\, $\mu_3\pm
\sigma_{\mu_3}$,\, $\mu_4\pm \sigma_{\mu_4}$,\, $\mu_5\pm
\sigma_{\mu_5}$, for the five indicators, respectively. Then to
calculate the best estimate $\mu$ for each GRB  event it is necessary to
make a weight-average of all available distance moduli, which is
given by:

\begin{eqnarray}
\label{muto}
\mu = \frac{1}{w}\sum_i^5w_i\mu_i \; ,
\end{eqnarray}

where $w_i=1/\sigma^2_i \quad \textrm{and}\quad  w=\sum_i^5w_i$

Then applying the error propagation law (see Eq.\ref{lpe}) to 
Eq.(\ref{muto}) we obtain the standard deviation associate
to this best estimate as $\sigma_{\mu}=1/\sqrt{w}$.

\subsection{Hubble Diagram}

In Table-\ref{datillos} we collect the distance moduli calculated for
each burst. Columns (1) and (2) give the GRB event and their redshift
($z$), Column (3)-(7) give the distance modulus $\mu_i\pm \sg_{\mu_i}$
calculated for each luminosity relation for the Cardassian cosmology:
$\omm=0.27$, $n=0.2$ and $H_0=72$ \phu\, , and the Column (8) gives the
combined distance modulus values obtained with the Eq.(\ref{muto})
and their uncertainties.

\begingroup
 \squeezetable
 \begin{table*}
 \caption{\label{datillos}Derived Distance Moduli}
 \begin{ruledtabular}
 \begin{tabular}{c c c c c c c c}
&     & $\mu_1$        &    $\mu_2 $    &  $\mu_3 $      &   $\mu_4$               &  $\mu_5 $      &   $\mu$     \\
 GRB  & $z$ &   (mag)        &     (mag)       &    (mag)       &   (mag)                & (mag)
&    (mag)    \\
 (1)  & (2)  &   (3)          &      (4)       &    (5)         &    (6)                 &  (7)
&     (8)       \\
 \\
 \hline
 \\
  970228\bd\bd  &  0.70 &      \cd       &  42.41\sr  1.23 &  42.31\sr  1.19 &       \cd       &  43.29\sr  1.20 &  42.67\sr  0.70\\
  970508\bd\bd  &  0.84 & 43.00\sr  1.14 &  42.99\sr  1.26 &  45.54\sr  1.05 &  43.78\sr  0.66 &  42.94\sr  1.21 &  43.77\sr  0.43\\
  970828\bd\bd  &  0.96 &      \cd       &  42.85\sr  1.21 &  43.96\sr  1.05 &  43.47\sr  0.66 &  43.14\sr  1.24 &  43.43\sr  0.47\\
  971214\bd\bd  &  3.42 & 48.63\sr  1.46 &  48.54\sr  1.22 &  47.44\sr  1.04 &       \cd       &  49.17\sr  1.32 &  48.31\sr  0.62\\
  980613\bd\bd  &  1.10 &      \cd       &       \cd       &  45.75\sr  1.35 &       \cd       &       \cd       &  45.75\sr  1.35\\
  980703\bd\bd  &  0.97 & 44.42\sr  0.97 &  44.81\sr  1.20 &  45.98\sr  1.04 &  43.45\sr  0.61 &  42.00\sr  1.25 &  44.02\sr  0.41\\
  990123\bd\bd  &  1.61 & 43.14\sr  0.95 &  44.69\sr  1.20 &  45.49\sr  1.05 &  45.38\sr  0.69 &       \cd       &  44.78\sr  0.46\\
  990506\bd\bd  &  1.31 & 44.69\sr  1.09 &  44.08\sr  1.19 &  44.07\sr  1.04 &       \cd       &  43.80\sr  1.21 &  44.18\sr  0.56\\
  990510\bd\bd  &  1.62 & 46.45\sr  1.02 &  45.11\sr  1.19 &  44.13\sr  1.03 &  45.38\sr  0.60 &  45.53\sr  1.20 &  45.34\sr  0.41\\
  990705\bd\bd  &  0.84 &      \cd       &  45.11\sr  1.20 &  43.47\sr  1.03 &  42.85\sr  0.66 &  45.66\sr  1.30 &  43.69\sr  0.47\\
  990712\bd\bd  &  0.43 &      \cd       &       \cd       &  41.75\sr  1.07 &  41.41\sr  0.69 &       \cd       &  41.51\sr  0.58\\
  991208\bd\bd  &  0.71 &      \cd       &  40.38\sr  1.22 &  42.09\sr  1.04 &       \cd       &  41.88\sr  1.20 &  41.52\sr  0.66\\
  991216\bd\bd  &  1.02 & 43.43\sr  1.01 &  42.38\sr  1.19 &  42.61\sr  1.04 &  43.53\sr  0.68 &  43.19\sr  1.23 &  43.16\sr  0.43\\
  000131\bd\bd  &  4.50 &      \cd       &  46.96\sr  1.22 &  47.59\sr  1.04 &       \cd       &  48.34\sr  1.37 &  47.58\sr  0.69\\
  000210\bd\bd  &  0.85 &      \cd       &  40.78\sr  1.22 &  43.68\sr  1.03 &       \cd       &  41.81\sr  1.21 &  42.28\sr  0.66\\
  000911\bd\bd  &  1.06 &      \cd       &  44.39\sr  1.21 &  45.54\sr  1.06 &       \cd       &  44.66\sr  1.31 &  44.94\sr  0.68\\
  000926\bd\bd  &  2.07 &      \cd       &  46.12\sr  1.22 &  44.13\sr  1.03 &       \cd       &  47.23\sr  1.44 &  45.49\sr  0.69\\
  010222\bd\bd  &  1.48 &      \cd       &  43.20\sr  1.19 &  43.56\sr  1.03 &  44.90\sr  0.57 &  43.55\sr  1.23 &  44.28\sr  0.43\\
  010921\bd\bd  &  0.45 & 42.76\sr  1.02 &  40.86\sr  2.42 &  43.07\sr  1.07 &       \cd       &  41.06\sr  1.26 &  42.34\sr  0.62\\
  012111\bd\bd  &  2.14 &      \cd       &      \cd        &  46.96\sr  1.06 &  44.98\sr  0.67 &       \cd       &  45.55\sr  0.57\\
  020124\bd\bd  &  3.20 & 47.73\sr  1.18 &  48.37\sr  1.28 &  46.14\sr  1.07 &  46.39\sr  0.67 &  47.22\sr  1.23 &  46.88\sr  0.44\\
  020405\bd\bd  &  0.70 &      \cd       &  43.90\sr  1.19 &  44.40\sr  1.12 &  43.41\sr  0.79 &  42.56\sr  1.21 &  43.55\sr  0.52\\
  020813\bd\bd  &  1.25 & 44.31\sr  0.97 &  45.19\sr  1.19 &  43.91\sr  1.04 &  43.90\sr  0.64 &  42.86\sr  1.21 &  44.01\sr  0.41\\
  020903\bd\bd  &  0.25 &      \cd       &       \cd       &  40.65\sr  1.29 &       \cd       &       \cd       &  40.65\sr  1.29\\
  021004\bd\bd  &  2.32 & 46.34\sr  1.21 &  46.60\sr  2.76 &  46.62\sr  1.18 &  45.71\sr  0.84 &  47.53\sr  1.34 &  46.34\sr  0.53\\
  021211\bd\bd  &  1.01 & 43.98\sr  0.94 &       \cd       &  42.19\sr  1.06 &       \cd       &  44.45\sr  1.21 &  43.51\sr  0.61\\
  030115\bd\bd  &  2.50 & 46.48\sr  1.09 &  47.25\sr  1.78 &  46.42\sr  1.14 &       \cd       &  45.36\sr  1.29 &  46.29\sr  0.63\\
  030226\bd\bd  &  1.98 & 46.84\sr  1.44 &  47.07\sr  1.97 &  46.64\sr  1.09 &  46.10\sr  0.69 &  46.35\sr  1.26 &  46.38\sr  0.48\\
  030323\bd\bd  &  3.37 &      \cd       &       \cd       &  46.74\sr  1.58 &       \cd       &  47.22\sr  1.46 &  47.00\sr  1.07\\
  030328\bd\bd  &  1.52 & 45.13\sr  1.43 &  44.61\sr  1.22 &  44.84\sr  1.04 &  44.49\sr  0.64 &       \cd       &  44.65\sr  0.47\\
  030329\bd\bd  &  0.17 & 41.94\sr  0.98 &  41.55\sr  1.20 &  39.57\sr  1.03 &  38.81\sr  0.61 &  40.49\sr  1.21 &  39.97\sr  0.41\\
  030429\bd\bd  &  2.66 &      \cd       &  47.65\sr  2.33 &  45.44\sr  1.14 &  46.44\sr  0.89 &  46.57\sr  1.26 &  46.28\sr  0.59\\
  030528\bd\bd  &  0.78 & 42.75\sr  1.05 &  44.75\sr  2.08 &  44.20\sr  1.10 &       \cd       &  46.08\sr  1.26 &  44.19\sr  0.62\\
  040924\bd\bd  &  0.86 & 43.83\sr  0.95 &       \cd       &  42.60\sr  1.04 &       \cd       &  45.09\sr  1.20 &  43.74\sr  0.61\\
  041006\footnotemark[2]\bd\bd &0.71 &  \cd & 44.09\sr  1.19 &42.38\sr  1.10 & 44.08\sr  0.73 &  43.26\sr  1.24 &  43.60\sr  0.50\\
  050126\bd\bd  &  1.29 & 45.37\sr  0.98 &  46.74\sr  1.41 &  45.77\sr  1.08 &       \cd       &  44.69\sr  1.27 &  45.57\sr  0.57\\
  050318\bd\bd  &  1.44 &      \cd       &  46.33\sr  1.22 &  44.20\sr  1.08 &  45.78\sr  0.72 &  46.14\sr  1.21 &  45.60\sr  0.49\\
  050319\bd\bd  &  3.24 &      \cd       &  46.49\sr  1.93 &       \cd       &       \cd       &  48.65\sr  1.23 &  48.03\sr  1.04\\
  050401\bd\bd  &  2.90 & 46.06\sr  1.15 &  46.94\sr  1.22 &  45.22\sr  1.06 &       \cd       &  48.56\sr  1.30 &  46.52\sr  0.59\\
  050406\bd\bd  &  2.44 & 48.16\sr  1.19 &       \cd       &  46.41\sr  1.43 &       \cd       &  48.96\sr  1.45 &  47.87\sr  0.77\\
  050408\bd\bd  &  1.24 & 45.17\sr  1.05 &       \cd       &       \cd       &       \cd       &  45.76\sr  1.27 &  45.40\sr  0.81\\
  050416\bd\bd  &  0.65 &      \cd       &       \cd       &  41.38\sr  1.12 &       \cd       &  45.22\sr  1.43 &  42.83\sr  0.88\\
  050502\bd\bd  &  3.79 & 47.26\sr  1.46 &  50.00\sr  1.30 &  46.89\sr  1.27 &       \cd       &  47.16\sr  1.39 &  47.87\sr  0.67\\
  050505\bd\bd  &  4.27 &      \cd       &  46.97\sr  1.59 &  46.86\sr  1.21 &  48.43\sr  0.99 &  47.60\sr  1.30 &  47.63\sr  0.61\\
  050525\bd\bd  &  0.61 & 44.01\sr  0.95 &  44.26\sr  1.19 &  41.93\sr  1.03 &  43.13\sr  0.71 &  43.32\sr  1.19 &  43.27\sr  0.43\\
  050603\bd\bd  &  2.82 & 45.69\sr  1.45 &  45.60\sr  1.23 &  45.48\sr  1.07 &       \cd       &  44.60\sr  1.20 &  45.32\sr  0.61\\
  050802\bd\bd  &  1.71 &      \cd       &  45.74\sr  2.52 &       \cd       &       \cd       &  45.34\sr  1.25 &  45.42\sr  1.12\\
  050820\bd\bd  &  2.61 & 45.87\sr  1.05 &       \cd       &  48.43\sr  1.08 &       \cd       &  44.97\sr  1.26 &  46.55\sr  0.65\\
  050824\bd\bd  &  0.83 &      \cd       &       \cd       &       \cd       &       \cd       &  43.22\sr  1.37 &  43.22\sr  1.37\\
  050904\bd\bd  &  6.29 &      \cd       &  47.06\sr  2.48 &  51.06\sr  1.13 &  49.19\sr  0.77 &  47.77\sr  1.27 &  49.27\sr  0.55\\
  050908\bd\bd  &  3.35 &      \cd       &       \cd       &  46.82\sr  1.06 &       \cd       &  46.91\sr  1.23 &  46.86\sr  0.80\\
  050922\bd\bd  &  2.20 & 46.46\sr  1.01 &  43.91\sr  1.25 &  45.86\sr  1.04 &       \cd       &  46.43\sr  1.21 &  45.77\sr  0.56\\
  051022\bd\bd  &  0.80 &      \cd       &  43.47\sr  1.19 &  44.69\sr  1.03 &  43.73\sr  0.57 &  43.33\sr  1.22 &  43.81\sr  0.43\\
  051109\bd\bd  &  2.35 &      \cd       &       \cd       &  46.59\sr  1.11 &       \cd       &  44.50\sr  1.27 &  45.69\sr  0.83\\
  050922\bd\bd  &  1.55 & 44.90\sr  0.96 &  44.64\sr  1.35 &       \cd       &       \cd       &  43.71\sr  1.30 &  44.52\sr  0.67\\
  060108\bd\bd  &  2.03 &      \cd       &  46.90\sr  3.83 &  46.87\sr  1.52 &       \cd       &  48.04\sr  1.74 &  47.34\sr  1.09\\
  060115\bd\bd  &  3.53 &      \cd       &       \cd       &  47.34\sr  1.04 &       \cd       &  48.38\sr  1.36 &  47.73\sr  0.83\\
  060116\bd\bd  &  6.60 &      \cd       &       \cd       &  49.29\sr  1.28 &       \cd       &  47.05\sr  1.42 &  48.29\sr  0.95\\
  060124\bd\bd  &  2.30 & 46.83\sr  1.09 &  47.39\sr  1.24 &  46.89\sr  1.10 &  46.96\sr  0.69 &  46.03\sr  1.27 &  46.87\sr  0.45\\
  060206\bd\bd  &  4.05 & 48.04\sr  1.44 &  45.90\sr  1.71 &  46.56\sr  1.06 &       \cd       &  45.71\sr  1.22 &  46.53\sr  0.65\\
  060210\bd\bd  &  3.91 & 47.48\sr  1.15 &  45.08\sr  1.27 &  47.52\sr  1.11 &  49.43\sr  0.73 &  46.63\sr  1.30 &  47.84\sr  0.46\\
  060223\bd\bd  &  4.41 & 47.47\sr  0.99 &  48.94\sr  1.48 &  47.39\sr  1.07 &       \cd       &  47.80\sr  1.23 &  47.74\sr  0.58\\
  060418\bd\bd  &  1.49 & 44.90\sr  0.96 &  45.19\sr  1.20 &  45.99\sr  1.04 &       \cd       &  45.24\sr  1.23 &  45.33\sr  0.54\\
  060502\bd\bd  &  1.51 & 43.60\sr  1.08 &  42.99\sr  3.53 &  46.81\sr  1.19 &       \cd       &  43.79\sr  1.31 &  44.65\sr  0.67\\
  060510\bd\bd  &  4.90 &      \cd       &  48.02\sr  1.77 &  48.89\sr  1.19 &       \cd       &       \cd       &  48.62\sr  0.99\\
  060526\bd\bd  &  3.21 & 48.22\sr  0.98 &  49.09\sr  1.38 &  44.88\sr  1.10 &  46.83\sr  0.82 &  48.53\sr  1.24 &  47.29\sr  0.47\\
  060604\bd\bd  &  2.68 & 45.16\sr  1.01 &       \cd       &  46.56\sr  1.07 &       \cd       &  47.98\sr  1.28 &  46.36\sr  0.64\\
  060605\bd\bd  &  3.80 & 45.14\sr  1.26 &       \cd       &  49.37\sr  1.19 &       \cd       &  46.44\sr  1.34 &  47.11\sr  0.73\\
  060607\bd\bd  &  3.08 & 45.08\sr  1.03 &  47.67\sr  1.31 &  47.56\sr  1.10 &       \cd       &  45.35\sr  1.25 &  46.33\sr  0.58\\
\end{tabular}
\end{ruledtabular}
\footnotetext[2]{\ Notice that in BS2007 this GRB has not
associated $\tlag$, but he obtained a $\mu_1$-value for this GRB
event.}
\end{table*}
\endgroup

\begin{figure}[t!]
\begin{center}
\includegraphics[scale=0.375]{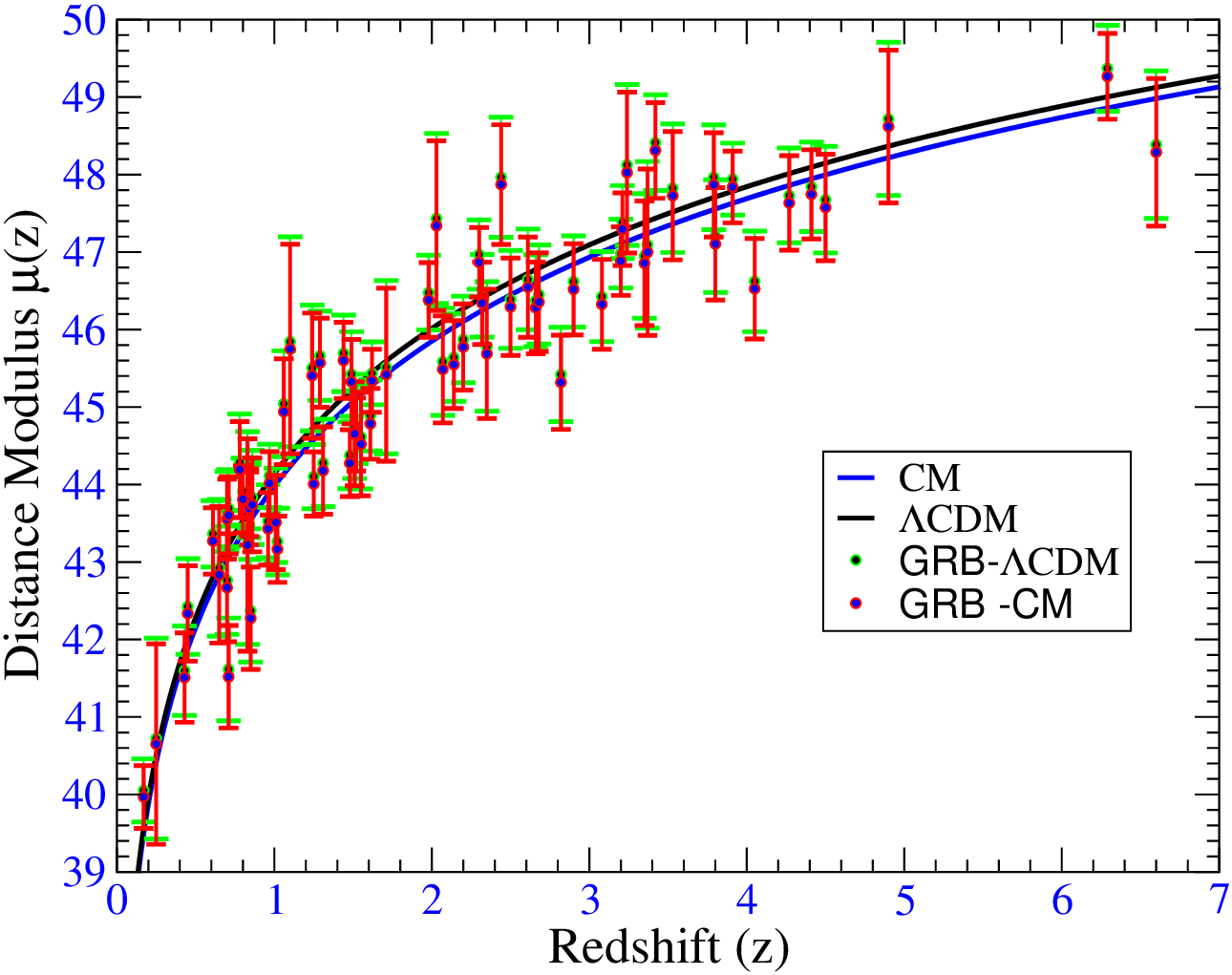}
\end{center}
\caption{\label{dh} Hubble Diagram of 69 GRBs after having been calibrated with 
Cardassian Cosmology. The blue line corresponds to the HD predicted by the 
Cardassian parameters used in this calibration; $\omm=0.27$, \,$n=0.2$. The 
black line corresponds to the HD for the Concordance Cosmology ($\Lambda$-CDM  
model with $\omm=0.27$, \,$w=-1$). Here we assumed $H_0 = 72$\,\phu .}
\end{figure}

\indent Fig.\ref{dh} shows the HD for the GRBs calibrated with the Cardassian 
Cosmology. Notice that the observational data points are consistent with the 
Cardassian model predictions because its $\chi$-square per degree-of-freedom 
$\chi^2_{dof} \approx 1.00-1.07$ (see next section).

\vskip 0.5truecm

\begingroup
\begin{table*}[!ht]
\caption{\label{chi2ana} Best-fit analysis }
\begin{ruledtabular}
\begin{tabular}{l c c c c}
Case\footnotemark[3]  &\om \sr \siom \footnotemark[4]      &  n \sr \si \footnotemark[4]       & $\chi^2$ &\chidof \\
          &                                     &                                   &         &        \\
 \hline
                       &                                     &                                   &        &    \\
exact \ho & 0.45 \sr 0.20 [$^{+0.14}_{-0.09}$]&-0.70 \sr 2.68 [$^{+0.00}_{-0.00}$]              &70.88& 1.06\\
exact \ho,prior\,\om & 0.28 \sr 0.04 [$^{+0.04}_{-0.04}$]& 0.27 \sr 0.16 [$^{+0.14}_{-0.19}$]    & 71.53  & 1.07\\
\prh                & 0.40 \sr 0.36 [$^{+0.25}_{-0.00}$]   & 0.26 \sr 1.66 [$^{+0.00}_{-0.00}$]  & 73.28  & 1.09\\
\prh+\rcmb          & 0.35 \sr 0.09 [$^{+0.11}_{-0.08}$]   & 0.39 \sr 0.22 [$^{+0.22}_{-0.23}$]  &  73.29 & 1.09\\
\prh.+\bao           & 0.33 \sr 0.04 [$^{+0.04}_{-0.04}$]   & 0.45 \sr 0.29 [$^{+0.26}_{-0.33}$] &  73.30 & 1.09\\
\prh.+\rcmb+\bao     & 0.32 \sr 0.04 [$^{+0.04}_{-0.04}$]   & 0.34 \sr 0.16 [$^{+0.14}_{-0.18}$] & 73.50  & 1.10\\
\end{tabular}
\end{ruledtabular}
\footnotetext[3]{Here we refer to prior a free parameter with Gaussian Prior distribution.
The marginalization procedure takes Hubble parameter $H_0$ as nuissance parameter.}
\footnotetext[4]{The parameter errors given in square brackets are
estimates obtained by the MINOS analysis of the base code MINUIT,
corresponding to one-standard deviation. In the case that they are
zero, the MINOS can not determine such errors.}
\end{table*}
 \endgroup

\section{\label{chi2analysis} Data Analysis: The Method}

We can now determine the best-fit to the set of cosmological
parameters $\te=(\omm,n)$ after imposing several constraints. 
Later on we combine GRBs data with those from the CMB and BAO 
to put tigh constraints on these parameters. To this end, we 
construct the log-likelihood function $\chi^2$ by assuming 
hypothetical $m$ and $M$ magnitudes fitted to a number of 
redshifts:


\begin{equation}
\chi^2_0(\te,H_0,b_i) = \sum_{i=1}^N \frac{[\mu_{te} (z_i,\te,H_0) - 
\mu_{\text{dat}}(z)(z_i,b_1, \ldots, b_5)]^2} {\sigma^2_{\mu_{\text{dat}}}(z_i)}\;  .
\label{chigrbs1}
\end{equation}

where $N$ is the number of data, $\mu_{\text{dat}}(z)$ and $\sigma_{\mu_{\text{dat}}}(z)$
are, respectively, the fitted distance modulus and its dispersion at redshift $z$ 
obtained from the calibration procedure. In addition, $\mu_{\text{th}}(z)$ is the 
theoretical prediction of the distance modulus and $b_i (i=1,\ldots,5)$ are the 
slopes obtained from the regression analysis.

We can obtain the function $\chi^2$ in the parameter space ($\te,H_0$) by 
marginalizing over the five nuisance parameters $b_i$. However, a reasonable 
approximation is to consider the values of the slopes fixed at the values
obtained in the regression analysis, because the GRB HD is nearly independent 
of the assumed cosmology. Therefore, the $\chi^2$ function reads

\begin{equation}
\chi^2_0(\te, H_0) = \sum_{i=1}^{N} \frac{(\mu_{\text{th}}(z_i,\te,H_0) 
- \mu_{\text{fit}}(z_i))^2} {\sigma^2_{\mu_{\text{fit}}}(z_i)}\; ,
\label{chiq}
\end{equation}

where $\mu_{\text{fit}}$ and $\sigma_{\mu_{\text{fit}}}$ are obtained in the 
previous section (see Eq.(\ref{muto}) and Column (8) in Table-\ref{datillos}).

\indent On the other hand, if we want to treat $\omm$ as a free parameter
with Gaussian prior distribution centered in $\omm^{\text{true}}$
with spread $\sg_{\omm-\text{prior}}$ \cite{goliath}, then we have
to use

\be
\chi^2_{\text{prior}}(\te,H_0) = \chi^2_0(\te,H_0) + \frac{(\omm -
\omm^{\text{true}})^2}{\sg^2_{\phantom{2}\omm-\text{prior}}}\; .
\label{chiprior}
\ee

\subsection{GRBs bounds $+$ constraints from other cosmological 
observations}

\indent To combine the GRBs data with the CMB we consider the
model-independent shift-parameter \rcmb, defined in terms of the
$H_0 $-independent luminosity distance: $D_L=H_0\,d_L$ (where
$d_L$ is the proper luminosity distance)\cite{rcmb}

\be
{\cal R} = \sqrt{\omm}\frac{D_L(z_r)}{1+z_r} = \sqrt{\omm}\int_0^{z_r}
\frac{dz}{E(z)} \; .
\label{sparama}
\ee

The shift-parameter ${\cal{R}}$ is related to the position of the
first acoustic peak of the CMB anisotropy power spectrum but it can
not be directly measured.  However, its value is estimated from the
data assuming a flat cosmology with dark matter and cosmological
constant. The shift-parameter is defined, approximately, as the
ratio of the sound horizon at recombination to the co-moving
distance to the last scattering. The quantity $E(z)$ is given by
Eq.(\ref{dlum1}) and $z_r$ = 1089 is the redshift of recombination.
(From the three-year results of WMAP \cite{spergel} Wang and
Mukherjee estimated \rcmb = 1.70 with spread $\sg_{\cal R}= 0.03$
\cite{wmuh}). Then, the statistical significance of a model is determined
evaluating $\chi^2_{\cal R} = ({\cal R} - 1.70)^2/\sg^2_{\cal R}$
together with the $\chi^2$ of the GRBs data, i. e.,

\be
 \chi^2 = \chi^2_0(\te,H_0) + \chi^2_{\cal R}(\te) \; .
 \label{chircmb}
\ee

\indent Meanwhile, to combine the GRBs data with the BAO we considered 
the \bao-parameter which describes the BAO peak in the matter perturbation 
power spectrum, given by \cite{eisen}:

\be {\cal A} = \sqrt{\omm} \left[\frac{1}{z_1\,E^{1/2}(z_1)}
\int_0^{z_1}\frac{dz'}{E(z')}\right]^{2/3} \; ,
\ee

where $z_1$ = 0.35 is the redshift at which the acoustic scale has been
found. Eisenstein \cite{eisen} found that the estimate value of the 
\bao-parameter is \bao = 0.469  with spread $\sg_{\cal A}=0.017$. 
Similarly to the shift-parameter, the statistical significance of a 
model is then determined evaluating $\chi^2_{\cal A}=({\cal A} - 
0.469)^2/ \sg^2_{\cal A}$ together with the $\chi^2$ of the GRBs data, 
i. e.,

 \be
 \chi^2=\chi^2_0(\te,H_0)+\chi^2_{\cal A}(\te) \; .
 \label{chibao}
\ee

This BAO peaks are clearly seen in the matter power spectrum, which
together with the overall shape put tight constraints on the model
parameters.

Therefore, if we combined the GRBs data plus CMB and BAO data the
most general expression to $\chi^2$ is given by: 

\be
\chi^2(\te,H_0)=\chi^2_0(\te,H_0)+\chi^2_{\cal
R}(\te)+\chi^2_{\cal A}(\te) \; .
\label{chigen} 
\ee

\indent To get the $\omm-n$ parameter space we use the
marginalization procedure choosing $H_0$ with Gaussian Prior
distribution as nuissance parameter. Therefore, we can define
a modified $\chi^2$-statistic by:

\begin{eqnarray}
\tilde{\chi}^2(\te ) = -2 \ln\left[\int_0^{\infty} dH_0 \;
\exp{[-0.5\chi^2 (\te,H_0)]} \pi(H_0)\right] \; ,
\label{chimarg}
\end{eqnarray}

 where $\quad \pi(H_0) = \frac{1}{\sqrt{2\pi\sg^2_{H_0}}}
\exp{\left(\frac{(H_0 - 72)^2} {2\, \sg^2_{\phantom{2}H_0}}
\right)}$. Here the function $\chi^2(\te,H_0)$ takes the form of
Eqs.(\ref{chiq}),\,(\ref{chircmb}),\,(\ref{chibao}) or (\ref{chigen})
depending on the analysis that will be performed. It is straightforward
to minimize  $\tilde{\chi}^2(\te)$  using the  base code MINUIT\cite{james}
or the function \emph{FindMinimum} of the Software Mathematica\cite{mathematica}
to find $\chi^2_{min}(\te)$, where $\chi^2_{min}$ is the minimum obtained for 
the best-fit parameters values $\bar{\te}=(\bar{\omm},\bar{n})$.

\indent The variable $\chi^2_{min}$ is random in the sense that it
depends on the random dataset that is used. Its probability distribution
is a $\chi^2$ distribution for $N-\nu$ degrees of freedom ($\nu$
is the number of free parameters). This implies that the 68\% of the
random parameters in a given dataset will give a $\chi^2$ such that

\be
\chi^2-\chi^2_{min}\leq \Delta\,\chi^2_{1\sigma}(\nu) \; ,
\label{cl1s}
\ee

where $\Delta\,\chi^2_{1\sigma}(\nu)$ is $2.3$ for two
($\nu=2$) free parameters. Thus, Eq.(\ref{cl1s}) defines the
1-$\sigma$  surface around the best-fit parameter. Similarly, it
can be shown that 95.4\% of the random numbers in the dataset will 
give a $\chi^2$ such that

\be
\chi^2-\chi^2_{min}\leq \Delta\,\chi^2_{2\sigma}(\nu) \; ,
\label{cl2s}
\ee

where $\Delta\,\chi^2_{2\sigma}(\nu)$ is 6.17 for two free parameters.
Thus, Eq.(\ref{cl2s}) defines the 2-$\sigma$ surface around the
best-fit and similarly for higher $\sigma$'s.

\indent The fixed values of $\Delta\,\chi^2_{i\sigma}$
($i=1,2,\ldots$) determine the boundary of the $i$-th confidence-level
contours and gives the probability of having a given $\chi^2$ for the
true values of the estimate parameters laying inside the boundary. This
criterion permit us to reject or accept any pair of ($\omm, n$)
parameter, and if $\chi^2/(N-\nu)\approx 1$ the fit is said to be good and
the data are said to be consistent with the considered cosmological model
under consideration. We divide the statistical analysis in two subsections,
considering the GRBs data alone and then combining GRBs with CMB and BAO data.

\section{Statistical analysis: Results}

\subsection{GRBs data alone}

In this subsection we consider only the GRBs dataset for the following 
three cases, and perform their statistical analysis and construct their
confidence-level contours, according to the method sketched above.



\begin{figure}[t!]
\begin{center}
\includegraphics[scale=0.4]{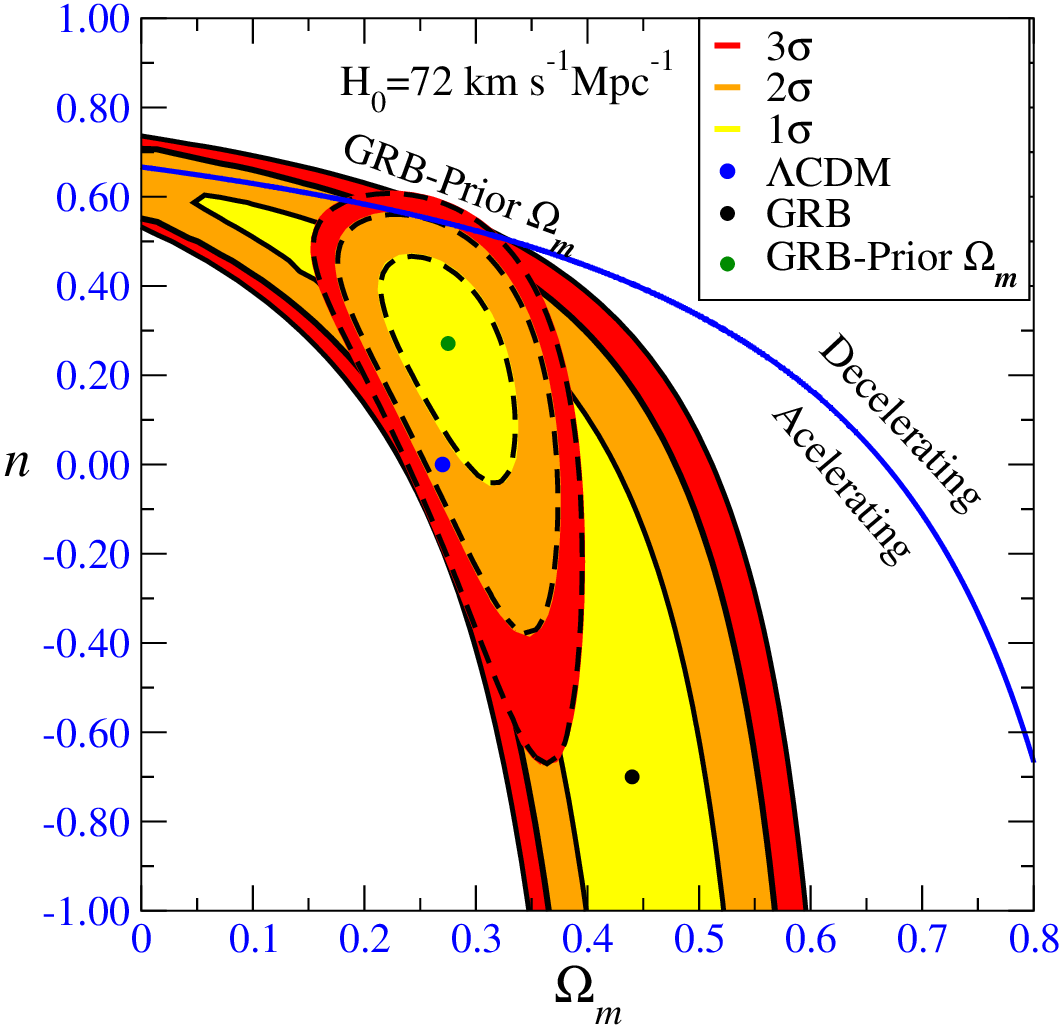}
\end{center}
\caption{\label{clcase2a} Confidence-level contours in the $\Omega_m-n$ parameter
space for the flat Cardassian Model. The solid lines separate the 1-$\sg$, 2-$\sg$ 
and 3-$\sg$ confidence-level (color) regions for the sole GRBs data, where the black 
point (0.45, -0.70) corresponds to the best fit. The encircling dashed lines divide 
the 1-$\sg$, 2-$\sg$ and 3-$\sg$ confidence-level regions obtained after imposing the 
prior knowledge on $\omm$, with the green point (0.28, 0.27) corresponding to the 
best-fit value, while the blue point (0.27, 0) corresponds to the cosmological constant 
$\Lambda$. One can see that in both cases the cosmological constant lies in the 2-$\sg$ 
level.}
\end{figure}

\subsubsection*{exact \ho}

Here we suppose an exact knowledge of the Hubble parameter today.
For the minimization procedure we use Eq.(\ref{chiq})
obtaining $\chi^2_{\text{min}}$ = 70.88 (\chidof$ =1.06$). In this
case, we have two free parameters, the best-fit value corresponds
to the point $(0.45, -0.7)$ in the $\omm-n$ parameter space. The
confidence contours are shown in the Fig.\ref{clcase2a}. Hence, we
see that the case $n=0$, which corresponds to the cosmological
constant $\Lambda$, lies in the 2-$\sg$ level.

\subsubsection*{exact \ho, prior \om}

In this case  we consider an exact knowledge of Hubble constant today 
and the matter density as free parameter with a uniform prior in the 
range $\omm = 0.27 \pm 0.04$. The best-fit analysis was performed using 
Eq.(\ref{chiprior}) obtaining $\chi^2_{\text{min}}$ = 71.53 (\chidof$ =
1.07$), and the values correspond to the point ($0.28, 0.27$) in the 
$\omm-n$ parameter space. The confidence-level contours correspond to 
the filled regions limited by dashed lines in Fig.\ref{clcase2a}. In 
this case, the point that corresponds to $\Lambda$ also lies in the 
2-$\sg$ level.

\subsubsection*{\prh}

\begin{figure}[t!]
\begin{center}
\includegraphics[scale=0.4]{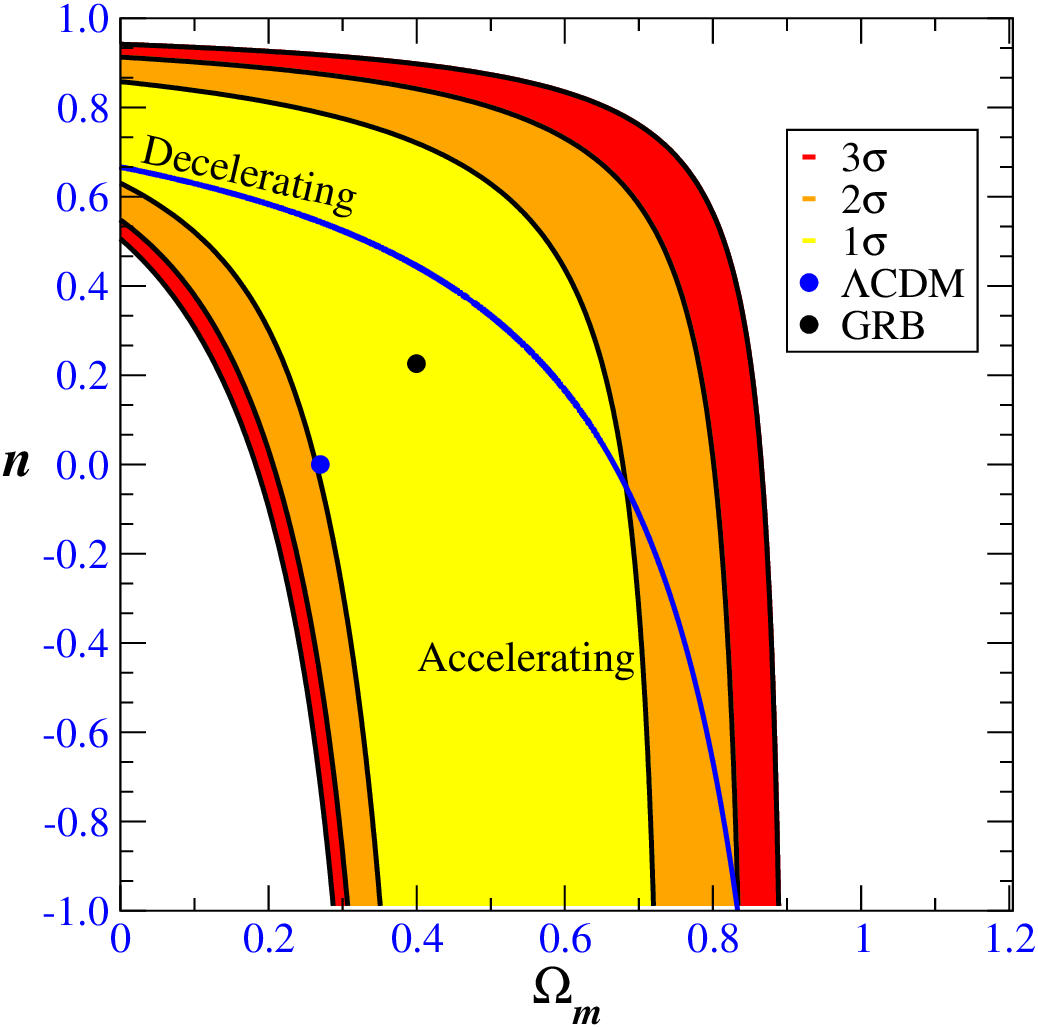}
\end{center}
\caption{\label{clcase4} Idem to Fig.\ref{clcase2a}, but now the black 
point (0.40, 0.26) corresponds the best-fit value, and the blue point 
(0.27, 0) corresponds to cosmological constant $\Lambda$. One can see 
that in this case the cosmological constant lies in the 1-$\sg$ level.}
\end{figure}

Here the best-fit analysis is the most realistic constraint for
GRBs. The minimization procedure was performed using the
Eqs.(\ref{chiq}) and (\ref{chimarg}) and the $\chi^2_{\text{min}}
= 73.28$, (\chidof$=1.09$). The best-fit value corresponds to the
point ($0.40, 0.26$) in the $\omm-n$  parmeter space. The result 
in the $\omm$-value agrees with the first one estimated by Schaefer
in BS2007.   The confidence-level contours correspond to the filled 
regions limited by solid lines in Figs.\ref{clcase4}, \,\ref{clcase67}, 
\,\ref{clcase68} and \ref{clcase69}. Hence, one can see that the 
cosmological constant lies in the 1-$\sg$ level.

\subsection{Combined Data}

In this subsection we analyze the role of the GRBs with combined
constraints from the CMB and BAO data.

\subsubsection*{\prh\, and\, \rcmb}

\begin{figure}[t!]
\begin{center}
\includegraphics[scale=0.4]{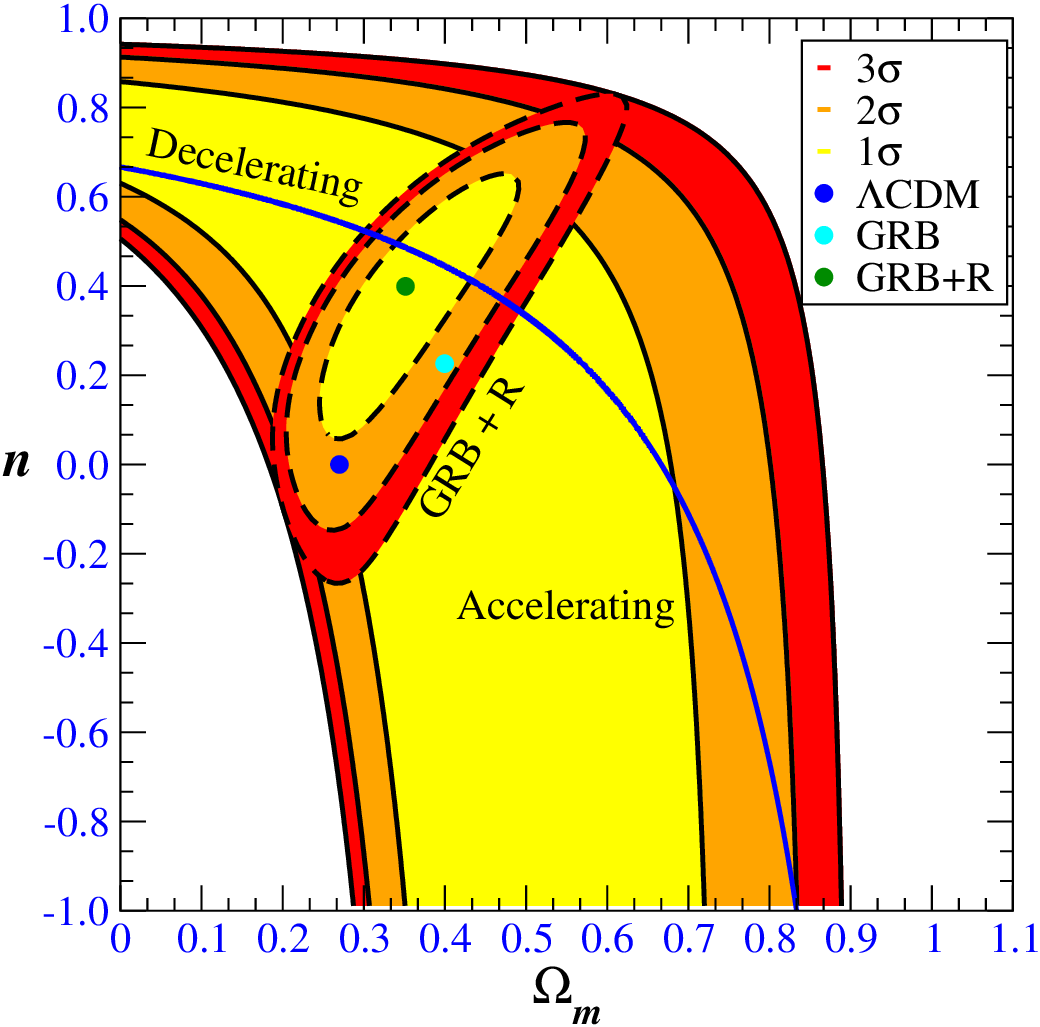}
\end{center}
\caption{\label{clcase67} Confidence-level contours in the $\omm-n$
parameter space  for the flat Cardassian Model with GRBs $+$ \rcmb. 
The solid lines separate the 1-$\sg$, 2-$\sg$ and 3-$\sg$ confidence-level 
(color) regions for the GRBs data alone, with best-fit value ($0.40, 0.26$) 
represented by the cyan point. The encircling dashed lines divide the 1-$\sg$, 
2-$\sg$ and 3-$\sg$  color-filled confidence regions of the bounds from 
GRBs + \rcmb, with the best-fit value ($0.35, 0.39$) represented by the green 
point. One can see that cosmological constant $\Lambda$ (blue point) lies in 
the 2-$\sg$ level.}
\end{figure}

Here we perform the minimization procedure using
Eq.(\ref{chimarg}) without $\chi^2_{\text{prior}}$ and
$\chi^2_{\cal A}$, and found $\chi^2_{\text{min}} = 73.29$,
(\chidof$=1.09$). The best-fit value corresponds to the point
($0.35, 0.39$), and the confidence contours are shown in the
Fig.\ref{clcase67}. One can see that the relevant constraint
on the parameter $n$, that would represent the dark energy 
and the cosmological constant $\Lambda$, lies in the 2-$\sg$ 
level.

\subsubsection*{\prh\, and \bao}

\begin{figure}[t!]
\begin{center}
\includegraphics[scale=0.4]{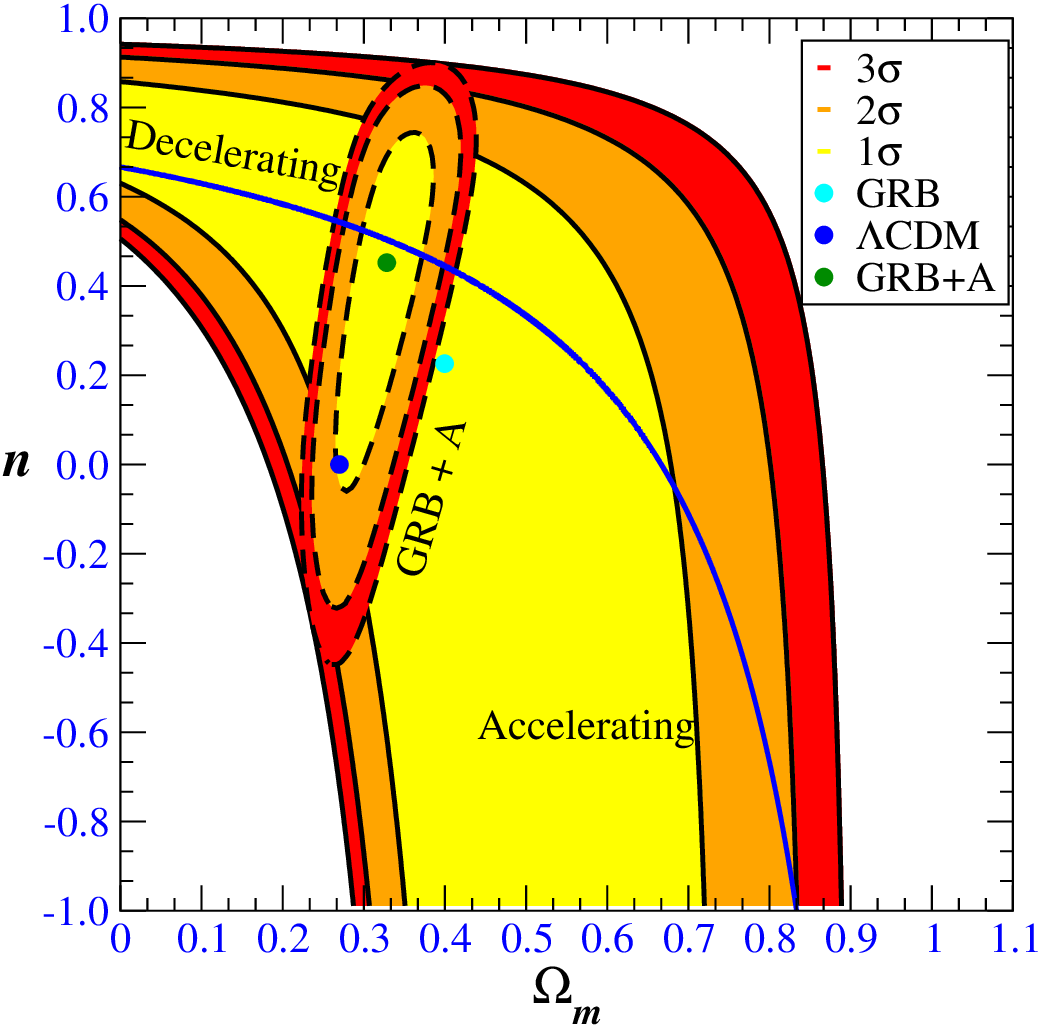}
\end{center}
\caption{\label{clcase68} Confidence-level contours in the $\omm-n$
parameter space for the flat Cardassian Model with GRB plus \bao. The 
solid lines separate the 1-$\sg$, 2-$\sg$ and 3-$\sg$ confidence-level 
(color) regions for the GRBs data alone, with best-fit value ($0.40, 
0.26$) represented by the cyan point. The encircling dashed lines divide 
the 1-$\sg$, 2-$\sg$ and 3-$\sg$  color-filled confidence regions of the 
bounds from GRBs plus \bao, with the best-fit value ($0.33, 0.45$) 
represented by the green point. One can see that cosmological constant 
$\Lambda$ (blue point) lies in the 1-$\sg$ level.}
\end{figure}

Here we perform the minimization procedure using
Eq.(\ref{chimarg}) without $\chi^2_{\text{prior}}$ and
$\chi^2_{\cal R}$, and found $\chi^2_{\text{min}} = 73.30$,
(\chidof$=1.09$). The best-fit value corresponds to the point
($0.33, 0.45$) in the $\omm-n$ parameter space, and the confidence
contours are shown in Fig.\ref{clcase68}. We can see that the
cosmological constant $\Lambda$ lies in the 1-$\sg$ level.

\subsubsection*{\prh,\, \rcmb\, and\,\bao}

\begin{figure}[t!]
\begin{center}
\includegraphics[scale=0.4]{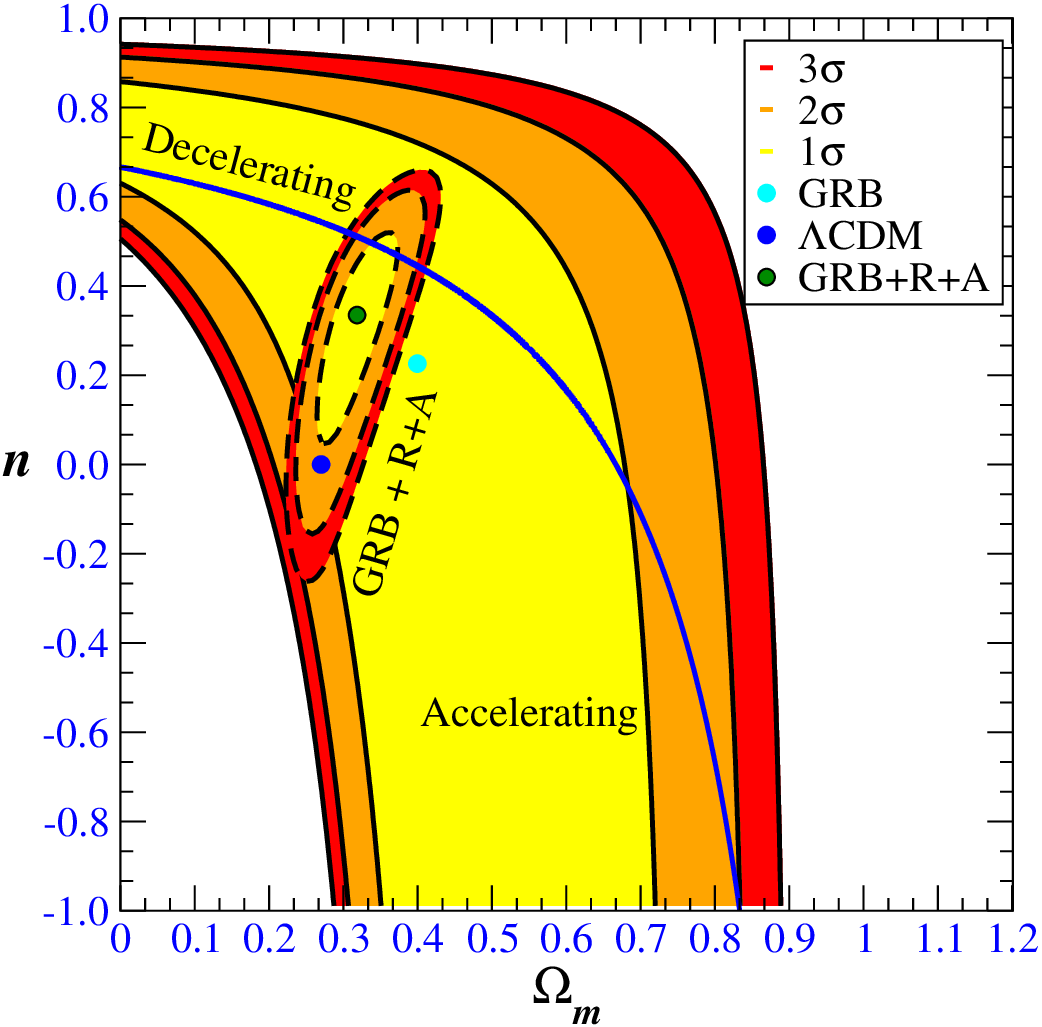}
\end{center}
\caption{\label{clcase69} Confidence-level contours in the \om-n
parameter space for the flat Cardassian Model with GRBs $+$ \rcmb\; 
$+$ \bao. The solid lines separate the 1-$\sg$, 2-$\sg$ and 3-$\sg$ 
confidence-level (color) regions for the GRBs data alone, with best-fit 
value ($0.40, 0.26$) represented by the cyan point. The encircling dashed 
lines divide the 1-$\sg$, 2-$\sg$ and 3-$\sg$  color-filled confidence 
regions of the bounds from the GRBs $\;+$ \rcmb $\;+$ \bao, with the  
best-fit value ($0.32, 0.34$) represented by the green point. Hence, 
one can see that cosmological constant $\Lambda$ (blue point) lies in 
the 2-$\sg$ level.}
\end{figure}

Here we consider the most general case for constraining the
cosmological parameters with the combined data. In this case, we
perform the minimization procedure using  Eq.(\ref{chimarg})
without $\chi^2_{\text{prior}}$, the $\chi^2_{\text{min}}$ value
is $73.50$ and (\chidof$=1.10$). The best-fit value corresponds to
the point ($0.32, 0.34$) in the $\omm-n$ parameter space, and the
confidence contours are shown in Fig.\ref{clcase69}. We can see
that the cosmological constant $\Lambda$ lies in the 2-$\sg$
level.

\noindent We collect our results for the several cases analyzed
above in Table-\ref{chi2ana}. One can see that our results are 
consistent with the result obtained by Bento et al. \cite{bento2005},
Gong and Duan \cite{gduan}, Zhu et al. \cite{zhu}. However in our 
analysis the cosmological constant lies in between 1-$\sg$ and 2-$\sg$.

\section{\label{dinamical} Other observational aspects}

Finally, we shall analyze the influence of the best-fit parameters on 
various cosmological parameters, including the Cardassian redshift of 
transition, which is one of the main observational features of that model.

\subsubsection*{Cardassian's redshift}

With the knowlegde of these best-fit parameters we calculate \zc\;  given by Eq.(\ref{zcarda}).
Notice that this redshift is the one where the universe enters the phase of accelerate expansion
driven by matter alone. This is the so-called Cardassian era. In general, it does not correspond
to transition era to the accelerate expansion. Our results for the first three cases  give $0.04 
< z_{\text{card}} < 0.56 $, in the most realistic case, where we considered the minimization 
procedure with the modified-$\chi^2$  for the GRBs data alone. For the combined data we obtain 
$0.41 < z_{\text{card}} < 0.55$. These results agree with the first initial values to $z_{\text{card}}$ 
given in the Table-I of the first Cardassian model proposed by Fresse and Lewis \cite{free}, which was 
obtained by using observations of SNIa and the CMB. For each case, the values obtained for this redshift 
are given in Column (2) in the Table-\ref{dinamic}.

\subsubsection*{Deceleration Parameter}

This parameter was introduced to characterize the dynamical
evolution of the universe based on the sign of the second time
derivative of the scale factor. It is defined as

\be
q = - \frac{\ddot{a}a}{\dot{a}^2} \; .
\label{qpar1}
\ee

The negative sign was put by-hand to come along with ancient ideas 
and ``observations'' suggesting that the universe was slowing down 
its expansion since the Big Bang. As we are interested in expressing 
it as a function of both the Hubble parameter and cosmological redshift 
we can easily rewrite Eq.(\ref{qpar1}) in the form:

\be
q(z) = \frac{1}{2}\frac{(1+z)}{E^2(z)} \frac{d\,E^2(z)}{dz}-1 \; .
\ee

\begin{figure*}[!htp]
\begin{center}
\subfigure[\label{qacel1} Deceleration parameter for GRBs data alone ]
{\includegraphics[scale=0.4]{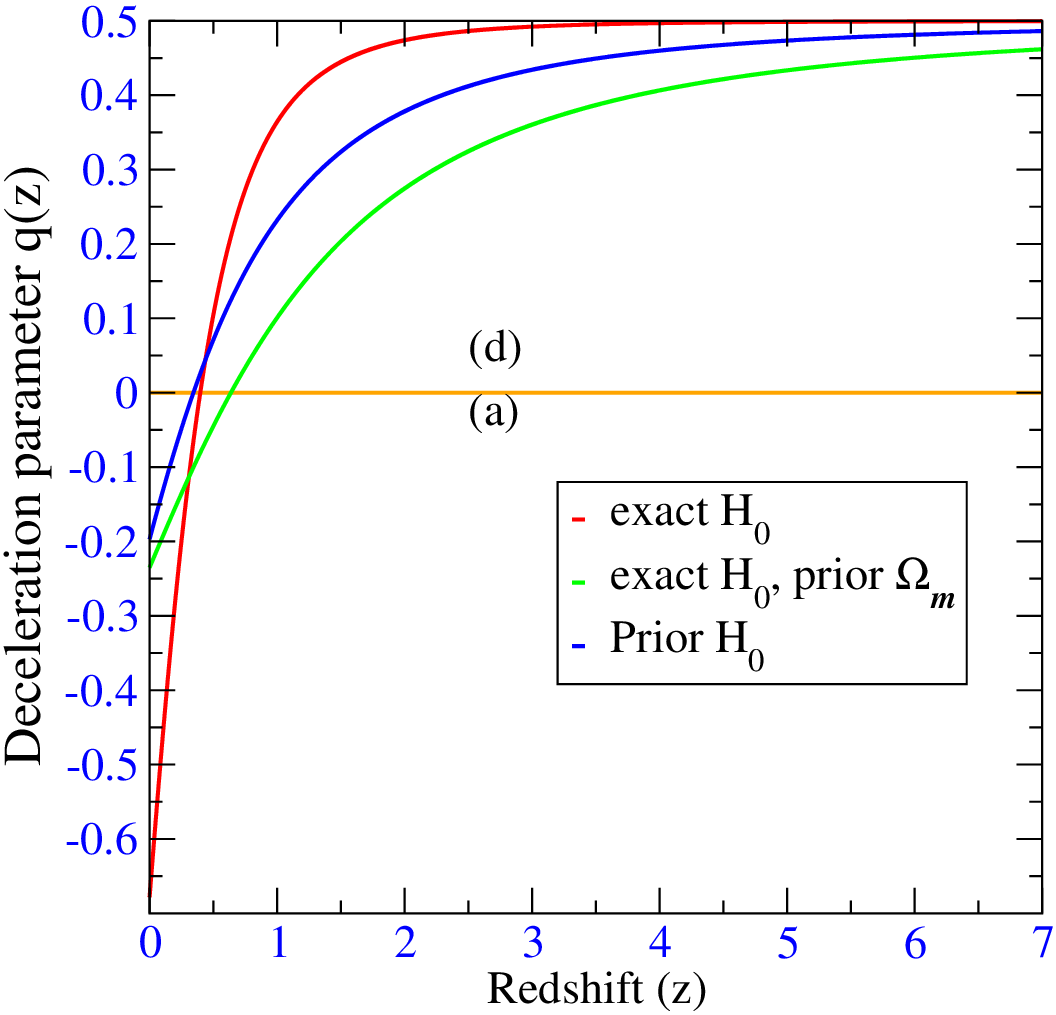}}
\hspace{1cm}
\subfigure[\label{qacel2} Deceleration parameter for combined GRBs + CMB data]
{\includegraphics[scale=0.4]{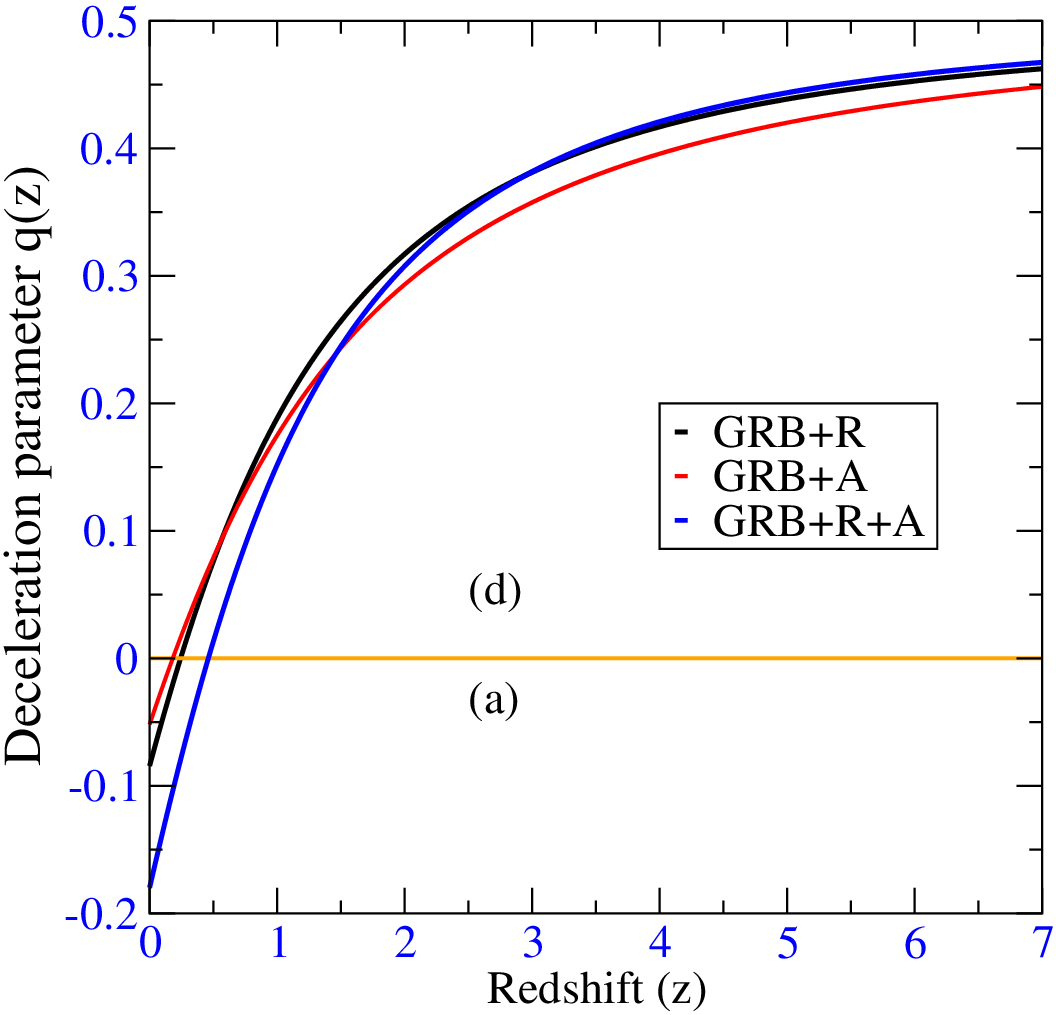}}
\end{center}
\caption{\label{qpard} Deceleration parameters for Cardassian model as
a function of the redshift for the values of best-fit solution. The
horizontal line, labeled (d)/(a) ($q(z)=0$), divide the region with
(a) acelerating or (d) decelerating expansion at a given redshift.
Clearly, we can see a transition from decelerating-to-accelerating
universe in the range $0.18 <z < 0.7$, which is approximately the
redshift range inferred from SNIa observations.}
\end{figure*}

It is clear that a negative value of this parameter stands for an
accelerating universe, and conversely, a positive value implies a
decelerating universe. For the Cardassian model we obtain

\begin{eqnarray}
 q(z) = \frac{1}{2} - \frac{3}{2} \frac{(1-n)\kappa(z)}{(1+\kappa(z))} \; ,
\end{eqnarray}

where $\kappa(z) = \left(\frac{\oxx}{\omm}\right)(1+z)^{-3(1-n)}$.
In Fig.\ref{qpard} we show  the deceleration parameter as a
function of the redshift for the values of the parameters obtained
with the best-fit analysis.

\indent The value $q_0 \equiv q(z=0)$ indicates the expansion rythm
today. In this case we obtain

 \begin{equation}
  q_0 = \frac{1}{2} - \frac{3}{2}(1-n) \oxx \; .
 \end{equation}

The values obtained for this parameter lay in the range $-0.91 < q_0 
< -0.05$ and are shown in Column (3) of Table-\ref{dinamic}.

\indent The transition redshift $z_{\text{acc}}$ can be obtained
by solving the equation $q(z)=0$. Our result agrees with the SNIa
observations $0.18 < z_{\text{acc}}< 0.68 $. The corresponding
values are shown in Column (4) of Table-\ref{dinamic}.

\subsubsection*{Age of universe}

Finally, we discuss the age-redshift relation in an alternatively
form which is independent of the Hubble parameter today. To this 
purpose we use the measurement of the quantity $H_0 \,t_0$:

\be
H_0 t_0 = \int_{0}^{\infty} \frac{dz}{(1+z)E(z)} \; ,
\label{agez}
\ee

where the dimensionless function $E(z)$ is obtained from
Eq.(\ref{dlum1}). Our result shows that $0.86 < H_0\,t_0 < 0.94$
(see Column (5) in Table-\ref{dinamic}). In order to verify our
result we have compared with several other estimates of the same
parameter. For instance, the  CMB alone contrains $t_0=13,7 \pm
0.25$ \edad \cite{spergel}, while the Globular Cluster Age imposes
$t_0=12.6^{3.4}_{-2.4}$ \cite{age2} for the age of universe. If we
also assume that  $H_0 = h (9.77813\edad)^{-1}$, with $h = 0.72
\pm 0.08$ from the HST Key Project, and supposing that the $H_0$
and $t_0$ measurements are uncorrelated, we obtain for the product
$H_0 \, t_0$ the following range: $0.79 < H_0\, t_0 < 1.27$ and
$0.67<H_0\,t_0<1.31$, respectively. Clearly, one can see that our
result is consistent with those estimates for the age of the
universe.

\begin{table}[!t]
\caption{\label{dinamic}Other observational properties}
\begin{ruledtabular}
\begin{tabular}{l c c c  c }
Case\footnotemark[3]    & \zc  & $q_0$  & \zac    &  $H_0t_0$\\
       (1)               &  (2) &  (3)   &  (4)    &   (5)   \\
 \hline
exact \ho                & 0.04 &  -0.91 & 0.37    &  0.90   \\
exact \ho, prior\,\om    & 0.56 &  -0.29 & 0.68    &  0.94   \\
\prh                     & 0.19 &  -0.20 & 0.34    &  0.88   \\
\prh  + \rcmb            & 0.41 &  -0.08 & 0.24    &  0.90   \\
\prh  + \bao             & 0.55 &  -0.05 & 0.18    &  0.86   \\
\prh  + \rcmb + \bao     & 0.46 &  -0.18 & 0.46    &  0.89   \\
\end{tabular}
\end{ruledtabular}
\footnotetext[3]{Here we refer to a prior as a free parameter with 
Gaussian Prior distribution.}
\end{table}

\section{\label{conclusiones}Discussion and Conclusions}

In the Cardassian model, the equivalent of the dark energy density 
arises from a modification of Friedmann's equations through the
introduction of a new function of the matter energy density in the 
form of an additional nonlinear term of mass. The outcoming model 
of the universe is then flat, matter dominated and accelerating.

Meanwhile, in order to understand the expansion history of the universe 
one needs to be able to trace it back in time up to very high redshifts
and construct its Hubble diagram from light-emitting cosmic sources laying
along the pathway to the Big Bang. In this perspective, GRBs offer a mean 
to extend up to redshifts $z>6$ the HD already built on SNIa observations $z<2$. 
GRBs are now known to have several light-curves (in different energy bands) 
and various spectral properties from which the luminosity of each burst can 
be calculated, once it is calibrated for a specific cosmology. This last 
procedure has been proved to be self-consistent \cite{bs2007}, and it turns 
GRBs useful standard candles for cosmographic studies.

Following the same procedure used by Schaefer in BS2007, we made a
simultaneous calibration  with the Cardassian Cosmology of those GRB 
events that had their redshifts properly estimated. This allows us
to construct the GRBs HD after using five empirical relations involving
luminosity distance indicators of these GRBs. The results of the
calibration procedure are shown in Figs.2-6. The Fig.1 states clearly 
that our method is self-consistent after comparing our results for the 
HD of GRBs calibrated with the concordance model with the one obtained
by Schaefer \cite{bs2007}. The HD for the GRBs calibrated with the 
Cardassian model and its comparison with the $\Lambda$-CDM is presented 
in Fig.7. We can see that the data of observed GRB events are consistent 
with both models. The small difference can be considered negligible if 
one takes into account that ($a$) the $\chi^2$ for both calculations are 
quite similar, and ($b$) the error bar of each GRB event is still quite 
large.

We have also performed a detailed statistical analysis and constructed 
the confidence-level contours from these GRBs data after imposing several
constraints on the cosmological parameters to be used for the analysis. 
First, considering exact knowledge of the Hubble parameter today, $H_0 = 
72$\phu\, , the best-fit value obtained corresponds to the point $(0.45, 
-0.7)$ in the $\omm-n$ parameter space. This point is shown in the associate 
confidence-level contour in Fig.8. The same figure shows the case where one 
considers also a uniform prior knowledge on the matter density ($\omm = 0.28
\pm 0.04$), which corresponds to the point $(0.28, 0.27)$. 
In Fig.9 we present the results of the analysis of the most specific case, that is, 
if one considers the lone GRBs dataset. We imposed a prior knowledge on the Hubble 
parameter today ($H_0=72 \pm 8$ \phu)\;, and the best-fit value obtained corresponds 
to the point ($0.40, 0.26$). This result also agrees with the first one obtained by 
Schaefer (2007).

We have also analyzed the bounds from GRBs when combined with constraints provided by
the CMB and BAO datasets. Considering the same prior knowledge on the Hubble parameter 
today, as we did before, the best-fit values obtained correspond to the points ($0.35, 
0.39$) in the case of GRBs + \rcmb \; (Fig.10), ($0.33, 0.45$) in the case of GRBs $+$ 
\bao \; (Fig.11), and ($0.32, 0.34$) in the case of GRBs $+$\rcmb $+$ \bao \; (Fig.12). 
The best-fit parameters obtained from this analysis are in agreement with those obtained 
from previous investigations that used SNIa data combined with CMB and BAO constraints.

Besides, with the knowledge of the best-fit parameters, we analyze their influence on 
other observational properties of this model. For the Cardassian redshift, our results 
for the first three cases (with fixed value of $H_0 $) give $0.04 < z_{\text{card}} < 
0.56 $. For the combined data we obtained $0.41 < z_{\text{card}} < 0.55$. These results 
agree with the very first initial values of $z_{\text{card}}$ calculated by Freese and 
Lewis \cite{free}, which was obtained by using observations of SNIa and the CMB. For the 
deceleration parameter, the values obtained lay in the range $-0.91 < q_0 < -0.05$. For the 
transition redshift $z_{\text{acc}}$ we obtained $0.18 < z_{\text{acc}}< 0.68$, which agrees 
with that inferred from SNIa observations. Finally, we obtained a couple of bounds for the 
age-redshift product $H_0 \, t_0$: ($a$) $0.79 < H_0\, t_0 < 1.27$ if one considers $t_0 = 
13.7 \pm 0.25$ \edad \; from CMB \cite{spergel}, and ($b$) $0.67 < H_0\, t_0 < 1.31$, if one 
considers $t_0 = 12.6^{3.4}_{-2.4}$ \edad \;  from the Globular Cluster Age \cite{age2}. Our 
results are consistent with those estimates for the age of the universe.

In summary, the GRBs HD has been built to demonstrate the reliability of GRBs as an 
independent tool to check the consistency of most current cosmological models over  
distance scales not allowed to SNIa. This diagram for GRBs shows a great  potential 
to make the GRBs observations a  complementary tool to SNIa,  large scale structure, 
baryon acoustic oscillations, an perhaps to CMB observations. In the near future, new 
studies based on these novel relations promise a major breakthrough in using GRBs to 
do precision cosmology. \newline


 \acknowledgements{H.D. would like to thank Martin Makler for several 
discussions on the statistical analysis and  help with the \emph{Mathematica
Software}. H.D. also thanks Prof. Orfeu Bertolami, and fellows
Cesar Castromonte and Gabriel Guerrer for assistance with the
implementation of the MINUIT base code. Fellows Rodrigo Turcati
and Jefferson Morais are also thanked for their help in
implementing the calibration procedure. Finally, H.D. thanks
fellow Guillermo Avenda\~no for assistance with the
construction of the contour region figures. H.J.M.C. and C.F.
thank Prof. R. Ruffini and the ICRANet Coordinating Centre,
Pescara, Italy, for hospitality during the final preparation of
this paper. The authors are also grateful to CAPES and CNPq for
the financial support received during the preparation of this
work.}

\end{document}